\documentclass[12pt]{article}
\usepackage{amsfonts,amssymb,epsfig,amsmath}
\usepackage{color}

\renewcommand{\baselinestretch}{1.2}

\def\Tr{{\rm Tr}}

\newcommand{\nn}{\nonumber}

\begin{document}

\makeatletter \@addtoreset{equation}{section} \makeatother
\renewcommand{\theequation}{\thesection.\arabic{equation}}
\renewcommand{\thefootnote}{\alph{footnote}}

\begin{titlepage}

\begin{center}
\hfill {\tt SNUTP15-002}\\
\hfill {\tt KIAS-15012}\\

\vspace{2cm}

{\Large\bf Little strings and T-duality}

\vspace{2cm}

\renewcommand{\thefootnote}{\alph{footnote}}

{\large Jungmin Kim$^1$, Seok Kim$^1$, Kimyeong Lee$^2$}

\vspace{0.7cm}

\textit{$^1$Department of Physics and Astronomy \& Center for
Theoretical Physics,\\
Seoul National University, Seoul 151-747, Korea.}\\

\vspace{0.2cm}

\textit{$^2$School of Physics, Korea Institute for Advanced Study,
Seoul 130-722, Korea.}\\

\vspace{0.7cm}

E-mails: {\tt kjmint82@gmail.com, skim@phya.snu.ac.kr,
klee@kias.re.kr}

\end{center}

\vspace{1cm}

\begin{abstract}

We study the 2d $\mathcal{N}=4$ gauge theory descriptions of little strings on
type II NS5-branes. The IIB strings on $N$ NS5-branes are described
by the $\mathcal{N}=(4,4)$ gauge theories, whose Higgs branch CFTs on $U(N)$
instanton moduli spaces are relevant. The IIA strings are described by
$\mathcal{N}=(4,4)$ $\hat{A}_{N-1}$ quiver theories, whose Coulomb branch CFTs are
relevant. We study new $\mathcal{N}=(0,4)$ quiver gauge theories for the IIA
strings, which make it easier to study some infrared observables. In particular,
we show that the elliptic genera of the IIA/IIB strings precisely map
to each other by T-duality.

\end{abstract}

\end{titlepage}

\renewcommand{\thefootnote}{\arabic{footnote}}

\setcounter{footnote}{0}

\renewcommand{\baselinestretch}{1}

\tableofcontents

\renewcommand{\baselinestretch}{1.2}
\numberwithin{equation}{section}

\section{Introduction}

Little string theories with $16$ supercharges \cite{Berkooz:1997cq,Seiberg:1997zk,
Aharony:1998ub,Giveon:1999zm,Aharony:1999ks} are obtained by decoupling limits of
the type II strings near $N$ NS5-branes. They are non-local theories without gravity.
Depending on whether we start from type IIA or IIB NS5-branes, the system has
$(2,0)$ or $(1,1)$ super-Poincare symmetry, respectively.
Since NS5-branes are one of the most difficult
nonperturbative objects to study in string theory, it would be very desirable to have
better understanding on these strings. Also, the type IIA little string theory
has interesting low energy limit given by interacting $(2,0)$ superconformal field
theories. Little strings have similarities with critical strings, and also
differences. The fact that it does not contain gravity would be the main difference,
with far-reaching implications.
However, being non-local theories, it inherits from the type II strings various
stringy properties, such as the T-duality. So after circle compactification,
the two little string theories are supposed to be T-dual to each other.

Unlike critical strings, noncritical little strings are difficult to study.
Some approaches to study them are: holographic approach
\cite{Aharony:1998ub,Giveon:1999zm}, discrete lightcone quantization
\cite{Berkooz:1997cq,Seiberg:1997zk,Witten:1997yu,Aharony:1999dw},
the double scaling limit \cite{Giveon:1999px,Giveon:1999tq}.
In particular, the DLCQ approach considers the little string theory compactified
on a small circle, in which one studies a sector with definite momentum which is
decoupled from the rest. Via T-duality, the DLCQ description can be obtained by
a large radius compactification of the T-dual strings with definite winding number.

In this paper, we study the QFTs living on the little strings macroscopically extended
on $\mathbb{R}^{1,1}$. They describe 2d decoupled degrees living on
these strings at low energy. Such theories are studied in detail
in the literature \cite{Witten:1997yu,Aharony:1999dw}. One starts from 2d
$\mathcal{N}=(4,4)$ gauge theories, which flow to interacting
CFTs and describe these strings. Compactifying these strings on large circles,
the ground energy is proportional to the radius times the winding
quantum number, much larger than the momentum energy scale. 
So we can consider a low energy decoupled
sector with fixed winding quantum numbers. They also have direct relevance to the study
of DLCQ little strings under T-duality, in which the momentum is fixed. In this paper,
we make a modest contribution to constructing such UV theories for the little
strings, on the IIA side starting from $\mathcal{N}=(0,4)$ gauge theories. The system
is proposed to flow to a CFT with enhanced $(4,4)$ SUSY. Compared to the $(4,4)$ gauge
theories discussed in \cite{Aharony:1999dw}, the new description has an advantage of 
manifestly having certain IR symmetries in UV, which is very crucial for computing some 
protected IR observables such as the elliptic genus. Our $(0,4)$ UV QFTs
are similar to those for the self-dual strings of the $(2,0)$
superconformal field theory, called `M-strings' \cite{Haghighat:2013gba}.

With T-duality, the spectrum of the circle compactified theories would be the same
for IIA and IIB little strings. We would like to probe this T-duality
with the above gauge theory descriptions for macroscopic strings. In general, these
descriptions are valid only when the compactification radii are large. As the
T-duality exchanges the IIA and IIB radii as $R_{A}=\frac{\alpha^\prime}{R_B}$,
the two gauge theory descriptions will never be simultaneously reliable.
However, one naturally expects that the protected BPS
spectrum would be reliable all the way to small radii.

In this paper, we study the T-duality of little strings in the BPS sector,
from the UV gauge theory descriptions. In particular, being
able to compute the elliptic genus indices on both IIA and IIB sides, thanks to our
new gauge theory descriptions, we can directly compare their BPS spectra. We find,
in fugacity expansions to highly nontrivial orders, that the two elliptic genera
precisely map to each other via T-duality.\footnote{In order to better define our spectral
problem, without continua coming from the `throat' regions
\cite{Witten:1997yu,Seiberg:1999xz,Aharony:1999dw}, we turn on the Fayet-Iliopoulos (FI)
term and the theta angle of the gauge theories on the worldsheet. Also, to avoid having
infrared problems with tensionless fractional strings or W-bosons, we separate the $N$
NS5-branes and study the massive spectra.} Apart from confirming the naturally expected
T-duality, our finding is establishing a very nontrivial identity between the elliptic
genera computed from the type IIA and IIB sides, so that alternative expressions can
be used to extract various properties which would have been very difficult to see
from the other viewpoints. For instance, we explain in section 5 how one can easily
understand the $SL(2,\mathbb{Z})\times SL(2,\mathbb{Z})$ transformation properties
of the elliptic genus, for the complex structure and Kahler parameters of the torus,
by using our T-dual expressions for the elliptic genus.

The rest of this paper is organized as follows. In sections 2 and 3, we explain
the 2d gauge theory descriptions of the IIB and IIA little strings, respectively,
and study their elliptic genera. In section 4, we study the T-duality of the two
elliptic genera, as well as extended duality/triality properties. In section 5, we study
the $SL(2,\mathbb{Z})$ transformation properties of the elliptic genus in various
fugacities. Section 6 concludes with brief discussions.

\section{IIB little strings}

\subsection{A brief review}

We first consider the type IIB little strings, which are the type IIB fundamental
strings bound to the NS5-branes. At low energy, the world-volume description of IIB
NS5-branes is given by 6d maximally supersymmetric Yang-Mills theory, with $(1,1)$
supersymmetry and $U(N)$ gauge group. The fields consist of the gauge field
$A_{\mu=0,\cdots,5}$, 4 scalar fields $\phi^{I=1,\cdots, 4}$, and fermions. These
degrees are provided by the D-strings ending on the NS5-branes.
The bosonic symmetry of the theory is $SO(1,5)\times SO(4)_{R}$. $SO(1,5)$ is the Lorentz
symmetry on the NS5-branes, and $SO(4)_{R}$ is the symmetry on their transverse directions, 
which rotates $\phi^{I}$. The Yang-Mills coupling constant is given by
\begin{equation}
 	g_{\rm YM}^2 = \frac{1}{T_{\rm NS5} (2 \pi \alpha')^{2}g^{2}_{s}} =(2\pi)^{3} {\alpha'}\;.
\end{equation}
Fundamental strings form threshold bounds with the NS5-branes. They are
identified as the instanton strings in the  6d SYM.
The instanton string tension is given by
\begin{equation}
	\frac{4 \pi^{2}}{g_{\rm YM}^2} = \frac{1}{2\pi \alpha'} = T_{\rm F1}\ ,
\end{equation}
agreeing with the tension of the fundamental string.
The coupling constant is independent of the 10d string coupling constant, $g_{s}$.
So one can take the little string theory limit, in which we take $g_{s} \rightarrow 0$
with fixed $\alpha'$. All the gravitational degrees of freedom are decoupled.

We shall consider $k$ macroscopically extended little
strings, extended along $\mathbb{R}^{1,1}$ part of $\mathbb{R}^{5,1}$. We are interested
in the dynamics of the degrees of freedom supported on these macroscopic strings,
decoupled from the rest of the 6d degrees at low energy. The system of $k$ F1 and $N$
NS5-branes admit a UV gauge theory description given by a $U(k)$ gauge theory with
$\mathcal{N}=(4,4)$ supersymmetry. The field theory is identical to that living on
the D1-D5 system via S-duality, and has been studied extensively in the literature,
e.g. \cite{Witten:1997yu,Aharony:1999dw,Seiberg:1999xz}. This 2d theory at low energy
can also be regarded as the worldsheeet description of the instanton strings of the
6d SYM theory. The gauge theory has the $U(k)$ $\mathcal{N}=(4,4)$ vector multiplet,
an adjoint hypermultiplet, and $N$ fundamental hypermulitiplets which host $U(N)$
global symmetry. These fields are shown in  Table \ref{IIB-multiplet}, and more 
details about this theory is explained in Appendix \ref{Appendix:2d-Action}. For 
later convenience, we
also show the supermultiplet structure with respect to the right-chiral $(0,4)$ SUSY.
\begin{table}[t!]
\begin{center}
\begin{tabular}{| c | c || c | c | c |}
	\hline
	$\mathcal{N}=(4,4)$  & $\mathcal{N}=(0,4)$  & Fields & $U(k)$ & $U(N)$ \\
	\hline
	\hline
	vector & vector & $A_{\mu}, \bar{\lambda}^{A \dot{\alpha}}_{+}$ & adj & 1 \\
	\cline{2-5}
	& twisted hyper 	& $\varphi_{aA},\;\bar{\lambda}^{\dot{\alpha}}_{a-}$ & adj & 1 \\
	\hline
	hyper & hyper 	& $a_{\alpha \dot{\beta}},\;\lambda^{A}_{\alpha-}$ & adj & 1 \\
	\cline{2-5}
	& Fermi 	& $\lambda_{a \beta+}$ & adj & 1\\
	\hline
	hyper & hyper 	& $q_{\dot{\alpha}},\;\psi^{A}_{-}$ & $\bar{\bf k}$ & $ {\bf N}$ \\
	\cline{2-5}
	& Fermi   & $\psi_{a-}$ & $\bar{\bf k}$ & ${\bf N}$ \\
	\hline
\end{tabular}
\caption{$\mathcal{N}=(4,4)$ supermultiplets for $k$ IIB strings.}\label{IIB-multiplet}
\end{center}
\end{table}
The bosonic symmetry preserved by the strings is $SO(1,1)\times SO(4)\subset SO(1,5)$
times $SO(4)_R$, where the latter is inherited from the R-symmetry of the 6d theory.
For $SO(4) \sim SU(2)_{L1} \times SU(2)_{R1}$ and $SO(4)_{R} \sim SU(2)_{L2} \times SU(2)_{R2}$, we introduce the following doublet indices,
\begin{equation}
	SU(2)_{L1}\;\rightarrow \;\alpha,\;SU(2)_{R1}\;\rightarrow \;\dot{\alpha},\;SU(2)_{L2}\;\rightarrow \;a,\;SU(2)_{R2}\;\rightarrow \;A\;.
\end{equation}
The fields in Table \ref{IIB-multiplet} and appendix A are given with this
convention. The 6d $(1,1)$ supercharges
can be written as $Q_{a \alpha+}$, $Q_{\alpha-}^{A}$, $Q^{\dot{\alpha}}_{a+}$,
$Q^{A\dot{\alpha}}_{-}$, where $\pm$ denote 6d chirality. These supercharges
satisfy the reality conditions given by
\begin{equation}
	Q_{a \alpha+} = -\epsilon_{\alpha \beta}\epsilon_{a b}(Q_{b \beta+})^{\dag},\;Q_{\alpha-}^{A} = \epsilon_{\alpha \beta}\epsilon^{AB}(Q_{\beta-}^{B})^{\dag},\;Q_{a+}^{\dot\alpha} = -\epsilon^{\dot\alpha \dot\beta}\epsilon_{a b}(Q_{b+}^{\dot\beta})^{\dag},\;Q^{A \dot{\alpha}}_{-} = \epsilon^{\dot\alpha \dot\beta}\epsilon^{AB}(Q^{B\dot{\beta}}_{-})^{\dag}.
\end{equation}
The strings extended on $\mathbb{R}^{1,1}$ preserve $Q^{\dot{\alpha}}_{a+}$ 
and $Q^{A \dot{\alpha}}_{-}$, forming 2d $\mathcal{N}=(4,4)$
supersymmetry. The $\pm$ subscripts on 2d fermions denote left/right chiralities, 
respectively.
In Table \ref{IIB-multiplet}, the $(4,4)$ Higgs branch fields
$a_{\alpha\dot\beta}$ and $q_{\dot\alpha}$ form the so-called ADHM data of $k$
multi-instantons of $U(N)$ gauge theory. This is because the IR dynamics of this
gauge theory will be describing the 6d instanton strings, as we shall explain in
more detail now.

The infrared dynamics of this $(4,4)$ theory has been studied in \cite{Witten:1997yu}.
Its low energy dynamics is described by two decoupled $(4,4)$ conformal field theories.
One is the conformal field theory on the Higgs branch described by a nonlinear sigma
model on the Higgs branch target space, given by $k$ instanton moduli space. 
Another is the conformal field theory on
the Coulomb branch. For studying the type IIB little strings, the Higgs branch CFT
is of relevance. The Coulomb branch degrees $\varphi_{aA}$ represent the motion of
the strings moving away from the 5-branes.

There is a peculiar singularity in the region near $q_{\dot\alpha}=0$,
$a_{\alpha\dot\beta}=0$, where the Higgs branch classically meets the Coulomb branch
\cite{Witten:1997yu,Aharony:1999dw,Seiberg:1999xz}. Quantum mechanically, this
region forms a `throat,' which is responsible for a continuum in the CFT
spectrum. The CFT can be deformed by turning
on the $SU(2)_{R1}$ triplet of Fayet-Iliopoulos term $\zeta^I$ ($I=1,2,3$) and the
theta angle $\theta$, after which the last continuum disappears. 
In particular, the Higgs branch moduli space becomes regular,
and the Coulomb branch is no longer connected to the Higgs branch even classically.
We shall consider the little string spectrum with nonzero FI term, from the elliptic
genus index \cite{Witten:1986bf} of the gauge theory. (The continuum will be
completely lifted, not only by the FI-term but also with the Coulomb VEV of
the 6d SYM to remove the infrared continuum.) In particular, with nonzero FI parameter,
the elliptic genus will acquire contribution only from the Higgs branch CFT for the
IIB little strings, and not from the Coulomb branch CFT that we are not interested in.

\subsection{The elliptic genus of IIB little strings}

In this subsection we shall define and explain the elliptic genus of the gauge theory
compactified on circle, counting $\frac{1}{4}$-BPS states in the Coulomb phase of the
6d theory, which shall be further studied in sections 4 and 5. This is a supersymmetric 
partition function on a torus with complex structure $\tau$.
We choose a supercharge $Q = Q^{A=1, \dot{\alpha} = \dot{2}}$ and define its index, 
with $q\equiv e^{2\pi i\tau}$,
\begin{align}\label{IIB-elliptic}
	Z_{\rm inst}^{\rm IIB}(\alpha_{i},\epsilon_{\pm},m;q,w) =\Tr\Big[(-1)^{F} w^{k} q^{H_{L}}\bar{q}^{H_{R}}e^{ 2 \pi i \alpha_{i}\Pi_{i}}e^{2 \pi i\epsilon_{-}(2J_{1L})}e^{2 \pi im (2J_{2L})}e^{2 \pi i\epsilon_{+}(2J_{1R}+2J_{2R})}\Big]\;\ .
\end{align}
$P$ is the momentum on the string compactified on the circle, and $H$ is the energy,
in the unit of inverse-radius $R_B^{-1}$ of the circle. $2H_{L}=H+P$ and $2H_{R} = H-P$ are
defined as the leftmoving and rightmoving momentum, respectively.
$J_{L1,2}$ and $J_{R1,2}$ are the Cartans of $SU(2)_{L1,2}$ and $SU(2)_{R1,2}$.
Since $\{Q,Q^{\dag}\} = 2 H_{R}$ and $Q$ commutes with all the other factors in the trace, the index counts only the BPS states annihilated by $Q$ and $Q^{\dag}$, and it is independent of $\bar{q}$.
$\Pi_{i}$'s are the Cartans of $U(N)$. $\alpha_{i}$'s are the chemical potentials for electric charges, interpreted as the background gauge field $A_{5}= {\rm diag}(\{ \alpha_{i}\})$ along the spatial circle, breaking $U(N)$ to $U(1)^{N}$.
We also introduce the fugacity variable, $w$, counting the winding number $k$ of
the little strings. For a given $U(k)$ gauge theory, we fix $k$. The above index is
the grand partition function. We use the subscript `inst' (standing for instantons)
in the 6d SYM interpretation, as we have already explained that the gauge theory
index will acquire contributions only from the Higgs branch.

It is also useful to consider the full index of the type IIB little string theory,
compactified on a circle with large radius $R_B\gg(\alpha^\prime)^{\frac{1}{2}}$.
The index is defined in the same way as
(\ref{IIB-elliptic}), where the trace is taken over the whole BPS Hilbert space of
the 6d theory in the Coulomb phase. This is a BPS partition function on 
$\mathbb{R}^4\times T^2$. Apart from (\ref{IIB-elliptic}), one finds extra
contribution from the 6d perturbative SYM states, which are decoupled with the
winding strings at low energy. The full index thus factorizes as
\begin{equation}
	Z_{\rm IIB}(\alpha_{i},\epsilon_{\pm},m;q,w) =Z^{\rm IIB}_{\rm pert}(\alpha_{i},\epsilon_{\pm},m;q) Z^{\rm IIB}_{\rm inst}(\alpha_{i},\epsilon_{\pm},m;q,w).
\end{equation}
The 6d perturbative index, $Z^{\rm IIB}_{\rm pert}$, counts the modes which only
carry momenta along the circle. This can be computed as follows. The momentum along the circle preserves supercharges $Q_{a\alpha+}$ and $Q^{A \dot{\alpha}}_{-}$, and breaks $Q^{\dot\alpha}_{a+}$ and $Q^{A}_{\alpha-}$. The Goldstino zero modes coming from the broken SUSY generators contribute to the single particle index with the following factor,
\begin{equation}
	 2^{4}\sinh\frac{2\pi i(m + \epsilon_{+})}{2}\sinh\frac{2\pi i (m - \epsilon_{+})}{2}\sinh \frac{2\pi i\epsilon_{1}}{2}\sinh \frac{2\pi i\epsilon_{2}}{2}\;,
\end{equation}
and the bosonic zero modes on $\mathbb{R}^{4}$ provides the factor
\begin{equation}
	\frac{1}{2^{4}\sinh^{2} \frac{2\pi i \epsilon_{1}}{2}\sinh^{2} \frac{2\pi i \epsilon_{2}}{2}}
\end{equation}
where $\epsilon_\pm\equiv\frac{\epsilon_1\pm\epsilon_2}{2}$.
Therefore, the single particle index of the particle carrying KK momentum is given by \cite{Kim:2011mv},
\begin{equation}
	\label{I+}
	I_{+}(\epsilon_{\pm},m) = \frac{\sinh \frac{2\pi i(m + \epsilon_{+})}{2}\sinh\frac{2\pi i (m - \epsilon_{+})}{2}}{\sinh \frac{2\pi i \epsilon_{1}}{2}\sinh \frac{2\pi i \epsilon_{2}}{2}} \;.
\end{equation}
The single particle index of the 6d perturbative particles is given by
\begin{eqnarray}
  z_{\rm sp}&=&N I_{+}(\epsilon_{\pm},m)\cdot \sum_{n=1}^{\infty}q^{n} + I_{+}(\epsilon_{\pm},m) \cdot \left(\sum_{i> j}^{N}e^{2 \pi i(\alpha_{i}-\alpha_{j})} +\sum_{i\neq j}^{N}\sum_{n=1}^{\infty}e^{2 \pi i(\alpha_{i}-\alpha_{j})}q^{n}\right)
  \nonumber\\
  &=&I_{+}\sum_{i> j}^{N}e^{2 \pi i(\alpha_{i}-\alpha_{j})}+I_{+}\left(N+\sum_{i\neq j}^{N}e^{2 \pi i(\alpha_{i}-\alpha_{j})}\right)\frac{q}{1-q}\ .
\end{eqnarray}
From this, $Z^{\rm IIB}_{\rm pert}$ is given by
\begin{equation}
	Z^{\rm IIB}_{\rm pert}(\alpha_{i},\epsilon_{\pm},m;q) =
PE\Big[z_{\rm sp}(\alpha_{i},\epsilon_{\pm},m;q)\Big] =  \exp\left[\sum_{p=1}^{\infty} \frac{1}{p}z_{\rm sp}(p\alpha_{i},p\epsilon_{\pm},pm;q^{p})\right]\;.
\end{equation}

The contribution of the winding IIB little strings $Z^{\rm IIB}_{\rm inst}$ 
is given in terms of the elliptic genera $Z_k$ of the $k$ instanton strings by
\begin{equation}
	Z^{\rm IIB}_{\rm inst}(\alpha_{i},\epsilon_{\pm},m;q,w) =\sum_{k=0}^{\infty}w^{k}Z_{k}(\alpha_{i},\epsilon_{\pm},m;q)\;,
\end{equation}
where $Z_{k=0}\equiv 1$. $Z_k$ is given by the sum of the terms characterized by $N$-colored Young diagrams, $Y= \{Y_{1},Y_{2},\cdots,Y_{N}\}\;$. The sum of the numbers of the boxes $\sum_{i=1}^{N}|Y_{i}|$ is $k$. The elliptic genus is given by \cite{Flume:2002az,Bruzzo:2002xf}
\begin{equation}
	\label{Instanton}
 	Z_{k}(\alpha_{i},\epsilon_{\pm},m;q) = \sum_{Y :\sum_{i}|Y_{i}|=k}\prod_{i,j=1}^{N}\prod_{s \in Y_{i}}\frac{\theta_{1}\left(q;E_{ij}+m- \epsilon_{-}\right)\theta_{1}\left(q;E_{ij} - m - \epsilon_{-}\right)}{\theta_{1}\left(q;E_{ij}-\epsilon_{1}\right)\theta_{1}\left(q;E_{ij} + \epsilon_2\right)}\;,
\end{equation}
where
\begin{equation}
	E_{ij} = \alpha_{i}-\alpha_{j}-\epsilon_{1}h_{i}(s)+\epsilon_{2}v_{j}(s).
\end{equation}
`$s$' denotes a box in the Young diagram $Y_{i}$. $h_{i}(s)$ is the distance from the box `$s$' to the edge on the right side of $Y_{i}$ that one reaches by
moving horizontally. $v_{j}(s)$ is the distance from `$s$' to the edge on the bottom side of $Y_{j}$ that one reaches by moving vertically. See e.g. \cite{Kim:2011mv} for more details 
and illustrations. The expression (\ref{Instanton}) may be computed by the contour 
integration formula given in terms of the Jeffrey-Kirwan residues \cite{Gadde:2013dda,Benini:2013nda,Benini:2013xpa}, as explained in \cite{Hwang:2014uwa}.

\section{IIA little strings}

Type IIA NS5-branes realize 6d IIA little string theory, with $\mathcal{N}=(2,0)$ supersymmetry. The light degrees should be made of $B_{\mu \nu}$ world-volume tensor gauge field whose field strength is self-dual in 6d, and 5 scalars, $\phi^{I=1,2,3,4}$ and $\phi$, and fermions. $\phi^{I=1,2,3,4}$ parametrize the transverse $\mathbb{R}^{4}$ of type IIA string theory, and $\phi$ is a compact scalar parametrizing
the position of the 5-branes along the
M-theory circle. The little strings are type IIA fundamental strings bound to
the NS5-branes. In M-theory, type IIA fundamental strings uplift to M2-branes 
wrapping the M-theory circle. The limit $g_{s}\rightarrow 0$ with a fixed $\alpha'$ 
yields the $\mathcal{N}=(2,0)$ little string theory.

\begin{figure}[t!]
  \begin{center}
    \includegraphics[width=10cm]{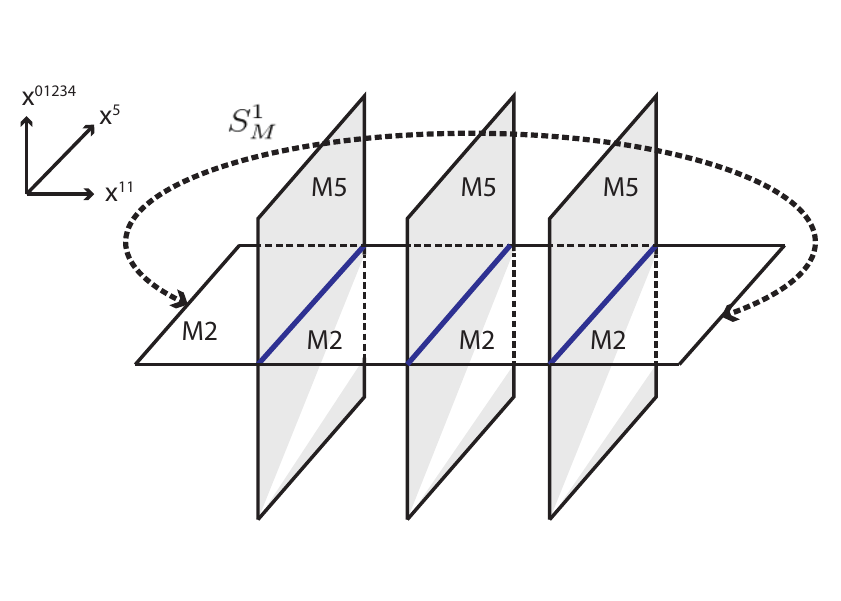}
\caption{M-theory brane uplift of the IIA little strings}\label{Fig:Brane-M}
  \end{center}
\end{figure}
\begin{table}[t!]
\begin{center}
\begin{tabular}{| c || c | c | c | c | c | c | c | c | c | c | c |}
	\hline
	& $x^{0}$ & $x^{1}$ & $x^{2}$ & $x^{3}$ & $x^{4}$ & $x^{5}$ & $x^{6}$ & $x^{7}$ & $x^{8}$ & $x^{9}$ & $x^{11}(S^{1}_{\rm M})$ \\
	\hline
	\hline
	$N$ M5 & $\times$ & $\times$ & $\times$  & $\times$ & $\times$ & $\times$ &&&&&  $\alpha_{i}$ \\
	\hline
	$n_{i}$ M2 & $\times$ &  &  &  &  & $\times$ &  &&&& $(\alpha_{i},\alpha_{i+1})$ \\
	\hline
\end{tabular}
\caption{M-theory brane uplift of IIA little strings.}\label{Brane-M}
\end{center}
\end{table}
The 2d little string gauge theory valid at $R_A\gg(\alpha^\prime)^{\frac{1}{2}}$
has been studied in \cite{Aharony:1999dw}, in the `Coulomb phase' with nonzero $\phi$,
separating all M5-branes along $x^{11}$. \cite{Aharony:1999dw} discussed it in
the context of type IIB strings on $A_{N-1}$ singularity, but let us review it in 
the M-theory context here.
The M-theory branes are shown in Fig. \ref{Fig:Brane-M}, where the M-theory circle
radius is given by $R_M=g_s\ell_s$ (where $\alpha^\prime=\ell_s^2$). See also
Table \ref{Brane-M} for coordinates. The tension
of the strings is given by $\sim\frac{R_M}{\ell_P^{3}}=\frac{R_M}{g_s\ell_s^3}
=\ell_s^{-2}$ in the original type IIA string theory, and we are interested in the
low energy 2d theory at excitation energy $E\ll g_{\rm YM}$, where $g_{\rm YM}$ is 
the 2d gauge coupling scale. To ease the
construction of this theory, we compactify $x^9$ direction along a circle with
radius $R_M^\prime$. Since NS5-branes are localized at $x^9=0$ and M2-branes are
attached to them, this compactification cannot be seen by the low energy CFT on
the strings, although it will be seen by the UV gauge theory we construct. Now
we make a 9-11 flip, regarding $x^9$ as the M-theory circle direction. 
The new type IIA theory would have its
own coupling and string scale $g_s^\prime$, $\ell_s^\prime$, satisfying
$g_s^\prime\ell_s^\prime=R_M^\prime$, $g_s^\prime(\ell_s^\prime)^3=\ell_P^3$.
The tension of the string given by the D2-branes suspended between NS5-branes
is $\frac{R_M}{g_s^\prime(\ell_s^\prime)^3}=\frac{R_M}{\ell_P^3}
=\ell_s^{-2}$, same as in the original type IIA picture. Now the low energy 2d
theory living on the D2-branes is easy to identify. It is a circular quiver
$U(k)^N$ gauge theory with $\mathcal{N}=(4,4)$ supersymmetry \cite{Douglas:1996sw}.
Each gauge node (labeled by $i=1,\cdots,N$) has vector multiplet fields
$A_\mu^{(i)}$, $a_{\alpha\dot\beta}^{(i)}$ and fermions, where $\alpha,\dot\beta$ are
the $SO(4)=SU(2)_{L1}\times SU(2)_{R1}$ spinor indices. There is a bi-fundamental
hypermultiplet mode connecting adjacent gauge node, and between $i$'th and $i+1$'th
node, the fields are denoted by complex scalars $\Phi^{(i)}_A$ and fermions. Compared
to be previous type IIB setting, or the original type IIA setting, in which we had
$SO(4)=SU(2)_{L2}\times SU(2)_{R2}$ R-symmetry, only the diagonal $SU(2)_D$ survives
after the $x^9$ circle compactification. So the doublet $A$ index 
can be regarded as the identification of the previous $a$ and $A$ indices. The $(4,4)$
supercharges are $Q^A_{\alpha+}$, $Q^A_{\dot\alpha-}$, subject to reality conditions. The 
$SU(2)_D$ UV symmetry is supposed enhance to full $SO(4)$ in IR, but is invisible in UV.
The incapability of seeing the second Cartan of $SO(4)$ from this UV theory will make it
impossible to study the full IR elliptic genus. This will be a motivation to 
study a $(0,4)$ supersymmetric UV gauge theory for the type IIA little strings, 
in section 3.1.

The coupling for the $i$'th $U(k)$ gauge field is given by
\begin{equation}\label{IIA-couplings}
  \frac{1}{g_{YM,i}^2}=\frac{(\alpha_{i+1}-\alpha_i)R_M\ell_s^\prime}{g_s^\prime}
  =\frac{(\alpha_{i+1}-\alpha_i)\ell_s^4}{(R_M^\prime)^2}\ ,
\end{equation}
which remains finite in the little string decoupling limit $g_s\rightarrow 0$.
All these couplings become large in the further IR limit on the strings
$E\ll g_{{\rm YM},i}$. One can turn on three FI parameters $\zeta_I^{(i)}$
for each $U(k)_i$ gauge group, which is a triplet of $SU(2)_D$ rotationg $678$.
This corresponds to the relative position of the $i+1$'th NS5-brane from the $i$'th
NS5-brane along $678$ directions. So one obtains the condition
$\sum_{i=1}^N\zeta_I^{(i)}=0$, since one should come back to the original NS5-brane
after going around the quiver once.

The gauge theory has $U(k)^N$ Coulomb branch, whose scalars represent the motion
of D2-branes along $1234$ directions. This would define the Coulomb branch CFT which
is relevant for studying the IIA little strings. On the other hand, the $N$ fractional
strings suspended between different adjacent pairs of NS5-branes can combine to make a
fully winding D2-brane along $x^{11}$, which may leave the NS5-brane along
the $6789$ directions (among which $x^9$ is the circle direction of the M-theory).
For instance, at $k=1$, the positions of the D2-branes along $678$ is parameterized
by the Higgs branch scalars, breaking $U(1)^N$ to $U(1)$ which lives on the D2-brane
separated from the NS5-branes. The $U(1)$ gauge field on this D2 would dualize to a
compact scalar, parametrizing the $x^{9}$ circle direction probed by the D2-brane.
More precisely, at $k=1$, the vanishing condition of the potential energy is given by
\begin{equation}
  \Phi_A^{(i)}a_{\alpha\dot\beta}^{(i)}-a_{\alpha\dot\beta}^{(i-1)}\Phi_A^{(i)}=0\ ,\ \
  \zeta^{(i)}_{I}+(\sigma_I)^A_{\ \ B}\Phi^{(i)}_A\bar\Phi^{B(i)}=
  (\sigma_I)^A_{\ \ B}\bar\Phi^{B(i-1)}\Phi^{(i-1)}_A\ .
\end{equation}
In the Higgs branch, one sets all $a^{(i)}_{\alpha\dot\beta}$'s to be equal,
so that the first equation is solved by breaking $U(1)^N\rightarrow U(1)$.
There is always a nonzero solution to the next equations, meaning that the Higgs
branch is always attached to the Coulomb branch.

Since the Higgs branch now represents the strings leaving the NS5-branes,
we are only interested in the Coulomb branch CFT in the IR limit.
However, the Higgs branch cannot be detached from the Coulomb branch CFT by
any deformation of the theory. This is in contrast to the 2d gauge
theories for the type IIB strings, in which case the Higgs branch CFT of our interest
could be detached from the Coulomb branch CFT  by turning
on $U(k)$ FI parameters. In fact, with generic FI term $\xi^{(i)}_I$, the Coulomb
branch will be all lifted by $U(1)^N\rightarrow U(1)$. Since the elliptic genus
formula of \cite{Benini:2013nda,Benini:2013xpa} is computing the index of CFT
with generic nonzero FI parameters, this formula will compute the unwanted Higgs
branch index, with lifted Coulomb branch. Apart from the absence of the $SU(2)_{L2}$
in UV, this is another reason that the above $(4,4)$ CFT is inconvenient for studying
the little string spectrum.

One can also add fractional D2-branes to this construction. Namely,
the number of $i$'th D2-branes between $i$'th and $i+1$'th NS5-branes can be
all different, $n_i$, forming a circular $U(n_1)\times\cdots\times U(n_N)$ quiver.

\subsection{$\mathcal{N}=(0,4)$ gauge theory descriptions}
\label{subsec:IIA little strings}

As explained, the $\mathcal{N}=(4,4)$ gauge theories for IIA little strings only see
$SU(2)_D\subset SO(4)$ part of the R-symmetry. Although we expect the symmetry enhancement
to happen in IR, this means that the UV gauge theory would be of limited use. 
Also, studying the spectrum of the Coulomb branch CFT will be difficult with the 
approaches of \cite{Benini:2013nda,Benini:2013xpa}. Closely following
the idea of \cite{Haghighat:2013gba,Haghighat:2013tka}, we shall engineer $(0,4)$
UV gauge theories for the IIA string systems which resolve all these problems.

Now on top of the IIA branes explained after the $x^9$-$x^{11}$ flip,
we also put one D6-brane extended along $012345,11$ and localized at $x^6=x^7=x^8=0$.
See Table \ref{Brane-IIA}. Now with a D6-brane uplifting to the Taub-NUT space 
in M-theory, the gauge theory $SU(2)$ which rotates $678$ directions in weakly coupled 
type IIA is interpreted differently in the IR CFT of this gauge theory. 
Namely, the low energy limit
of the 2d gauge theory is realized by taking the M-theory limit
$R_M^\prime\rightarrow\infty$ (after the $9$-$11$ flip): see (\ref{IIA-couplings}).
So the embedding of the UV gauge theory's symmetries into the infrared R-symmetry
has to be understood in the $R_M^\prime\rightarrow\infty$ limit, where we have
$\mathbb{R}^4$. The $SO(3)$ rotating the asymptotic $\mathbb{R}^3$ of Taub-NUT rotates
the $\mathbb{R}^4$ as $SU(2)_{R2}$ in `IR.' 
Also, after compactifying one more circle $x^{5}$,
we can turn on a background gauge field of the D6-branes, as $A^{i}_{5}+iA^{i}_{11}\equiv m^{i}\sim(m,2m,3m,\cdots,Nm)$ with nonzero $B_{5,11}$ turned on.
The parameter $m$ realizes the chemical potential for the Cartan of $SU(2)_{L2}$
\cite{Haghighat:2013gba,Haghighat:2013tka}. Thus, we can turn on full set of
$SO(4)_R$ chemical potentials of the partition function in this setting. From 
the 2d gauge theory viewpoint, adding one D6-brane just affects the way we connect
the UV regime $\mathbb{R}^3\times S^1$ at weak-coupling with the IR regime $\mathbb{R}^4$ at strong coupling. Since the IR brane configuration is complete the same as the original
M2-M5 system, we expect the $(0,4)$ gauge theory to flow to the same $(4,4)$ CFT on
the Coulomb branch. (However, see section 4 for discussions on irrelevant decoupled 
sectors within this gauge theory.)
\begin{table}[t!]
\begin{center}
\begin{tabular}{| c || c | c | c | c | c | c | c | c | c | c |}
	\hline
	& $x^{0}$ & $x^{1}$ & $x^{2}$ & $x^{3}$ & $x^{4}$ & $x^{5}$ & $x^{6}$  & $x^{7}$ & $x^{8}$ & $x^{11}(S^{1}_{\rm})$  \\
	\hline
	\hline
	$N$ NS5 & $\times$ & $\times$ &  $\times$ & $\times$ & $\times$ & $\times$ &
&&& $\alpha_{i}$\\
	\hline
	$n_{i}$ D2 & $\times$ &  &   &  &  &  $\times$ & &&& $(\alpha_{i},\alpha_{i+1})$ \\
	\hline
	$1$ D6 & $\times$ & $\times$ & $\times$  & $\times$ & $\times$ & $\times$ &  &&& $\times$ \\
	\hline
	%\hline
	%Taub-Nut & &  &  &  &  &  &  & $\times$ & $\times$ & $\times$ & $\times$ \\
	%\hline
\end{tabular}
\caption{Brane construction of 2d $\mathcal{N}=(0,4)$ gauge theory
}\label{Brane-IIA}
\end{center}
\end{table}

\begin{figure}[t!]
  \begin{center}
    \includegraphics[width=13cm]{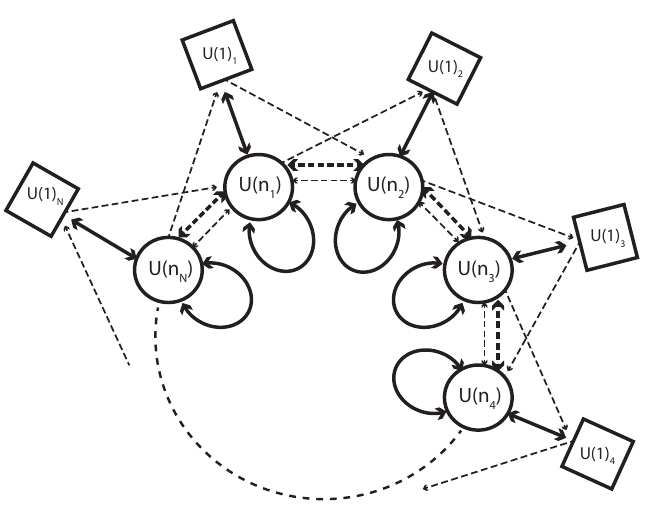}
\caption{$\hat{A}_{N-1}$ quiver diagram of the 2d $\mathcal{N}=(0,4)$ gauge theory for
the IIA strings. Solid lines denote the hypermultiplets, thin dashed lines denote the
Fermi multiplets, and thick dashed lines denote the twisted hyper multiplets. }\label{IIAquiver}
  \end{center}
\end{figure}

\begin{table}[t!]
\begin{center}
\begin{tabular}{| c || c | c | c |}
	\hline
	Multiplet & Fields & $U(n_{i})$ & $U(1)_m $ \\
	\hline
	\hline
	Vector & $A^{(i)}_{\mu}, \bar{\lambda}^{(i)A \dot{\alpha}}_{+}$ & adj$_i$ & 0 \\
	\hline
	Hyper 	& $q^{(i)}_{\dot{\alpha}},\;\psi^{(i)A}_{-}$ & ${\bf n}_i$ & 0 \\
	\hline
	Hyper 	& $a^{(i)}_{\alpha \dot{\beta}},\;\lambda^{(i)A}_{\alpha-}$ & 
    adj$_i$ & 0 \\
	\hline
	Twisted hyper 	& $\Phi^{(i)}_{A},\;\Psi^{(i)\dot{\alpha}}_{-}$ 
    & $({\bf n}_{i-1},\bar{\bf n}_{i})$ & $1$ \\
	\hline
	Fermi 	& $\Psi^{(i)}_{\beta+}$ & $({\bf n}_{i-1},\bar{\bf n}_{i})$ & $1$\\
	\hline
	Fermi   & $\psi^{(i)}_{+}$ & ${\bf n}_{i}$ & $1$ \\
	\hline
	Fermi   & $\tilde\psi^{(i)}_{+}$ & $\bar{\bf n}_{i}$ & $-1$ \\
	\hline
\end{tabular}
\caption{Fields of the $\mathcal{N}=(0,4)$ quiver gauge theory}\label{Multiplet-IIA}
\end{center}
\end{table}
A 2d $\mathcal{N}=(0,4)$ UV gauge theory is engineered from this brane setting,
with supercharges given by $Q^{A \dot\alpha}$. The fields can be characterized
again by a circular quiver of Fig. \ref{IIAquiver}. Each circular node involves $\mathcal{N}=(0,4)$ $U(n_{i})$ gauge multipletiplet $(A_{\mu}, \bar\lambda^{A \dot{\alpha}}_{+})$, and a $\mathcal{N}=(0,4)$ adjoint hypermultiplet $(a_{\alpha\dot{\beta}}, \lambda^{A}_{\alpha-})$, denoted by the solid lines. 
$n_{i}$'s are the number of the D2-branes suspended between adjacent NS5-branes. 
Thick dashed lines between two circular nodes denote the bi-fundamental twisted hypermultiplets $(\Phi_{ A}, \Psi^{\dot{\alpha}}_{-})$. Thin dashed lines between two circular nodes denote the bi-fundamental fermi multiplets $\Psi_{\beta+}$. 
D6-brane introduces extra fields: fundamental hyper multiplets 
$(q_{\dot{\alpha}},\psi^{A}_-)$ and Fermi multiplets $\psi_{+},\tilde\psi_+$. 
These fields are are summarized in Table. \ref{Multiplet-IIA}.
As explained in the previous paragraph, and just like \cite{Haghighat:2013gba}, the
chemical potentials for $U(1)_i$ and $U(1)_{i+1}$ are locked as $m_{i+1}-m_i=m$, 
so that one just has one $U(1)_m$. 
Compared to the previous $(4,4)$ gauge theory for the IIA strings,
the $(0,4)$ fields on the first and third lines of Table. \ref{Multiplet-IIA}
are forming the $(4,4)$ vector multiplet, which we decomposed as above since
the system does not preserve $(4,4)$ SUSY. Also, the fields on the fourth
and fifth lines form the previous $(4,4)$ hypermultiplet. They again make
a twisted Higgs branch, which represents the degrees of freedom of fully winding
D2-branes leaving the NS5-branes. The Coulomb branch of the $(4,4)$ theory is
replaced here by the Higgs branch formed by the second and third lines, which
is our main interest to study the IIA little strings.

The SUSY action of the $(0,4)$ gauge theory can also be easily
constructed. From the $(0,2)$ supersymmetric formalism, one has to determine the 
holomorphic potentials $E_\Psi$, $J_\Psi$ for each Fermi multiplet $\Psi$, ensuring 
the $(0,4)$ SUSY enhancement. For instance, see \cite{Tong:2014yna} for how this can 
be done. Here, following \cite{Tong:2014yna}, we simply write down these potentials
for our theory. Let us call the $(0,2)$ Fermi multiplet from the $(0,4)$ vector multiplet
as $\Lambda_i$, which is made of $\bar\lambda^{1\dot{1}}$ and $\bar\lambda^{2\dot{2}}$.
Then one should first take
\begin{equation}
  J_{\Lambda_i}=q_i\tilde{q}_i+[B_i,\tilde{B}_i]-\xi_{\mathbb{C}}\ ,\ \
  E_{\Lambda_i}=\Phi_{i+1}\tilde{\Phi}_{i+1}-\tilde\Phi_{i}\Phi_{i}\ ,
\end{equation}
for $(0,4)$ SUSY \cite{Tong:2014yna}. Here and below, we use the chiral superfield notation
$q_{\dot\alpha}=(q,\tilde{q}^{\dag})$, $a_{1\dot\beta}=(B,\tilde{B}^\dag)$,
$\Phi_A=(\Phi,\tilde\Phi^\dag)$ for a while. We also inserted the FI parameter
$\xi_{\mathbb{C}}$ for later use, which corresponds to turning on worldvolume
$B_{\mu\nu}$ field on $1234$ directions. The above $J,E$ should be accompanied
by other $J,E$ functions for other Fermi fields, to satisfy $\sum_\Psi E_\Psi J_\Psi=0$
after summing over all Fermi multiplets $\Psi$. This is another requirement from SUSY.
To meet the last condition, one should turn on the following potentials for other Fermi
multiplet fields:
\begin{eqnarray}
  &&E_{\tilde\psi_i}=\tilde{q}_{i-1}\Phi_{i}\ ,\ \ 
  J_{\tilde\psi_i}=-\tilde\Phi_i q_{i-1}\ ,\ \
  E_{\psi_i}=\Phi_{i+1} q_{i+1}\ ,\ \ J_{\psi_i}=\tilde{q}_{i+1}\tilde\Phi_{i+1}
  \nonumber\\
  &&E_{\Psi_i}=\Phi_iB_i-B_{i-1}\Phi_i\ ,\ \
  J_{\Psi_i}=\tilde{B}_i\tilde\Phi_i-\tilde\Phi_i\tilde{B}_{i-1}\ ,\ \ \nonumber\\
  &&E_{\tilde\Psi_i}=\tilde{B}_{i-1}\Phi_i-\Phi_i\tilde{B}_{i}\ ,\ \
  J_{\tilde\Psi_i}=B_{i}\tilde\Phi_i-\tilde\Phi_iB_{i-1}\ .
\end{eqnarray}
The bosonic potential is $V=\sum_{\Psi}(|J_\Psi|^2+|E_\Psi|^2)
+\frac{1}{2}\sum_iD_i^2$ with $D_i$ given by
\begin{equation}
  D_i=q_iq_i^\dag-\tilde{q}_i^\dag\tilde{q}_i+[B_i,B_i^\dag]
  +[\tilde{B}_i,\tilde{B}_i^\dag]-\Phi_i^\dag\Phi_i+\tilde\Phi_i\tilde\Phi_i^\dag
  +\Phi_{i+1}\Phi_{i+1}^\dag-\tilde\Phi_{i+1}^\dag\tilde\Phi_{i+1}-\xi_{\mathbb{R}}\ .
\end{equation}
After some rearrangement, one obtains
\begin{eqnarray}
  V&=&\frac{1}{2}\sum_{i=1}^N\left[\frac{}{}\!\right.
  \left(q_{i\alpha}(\sigma^m)^{\dot\alpha}_{\ \ \dot\beta}\bar{q}_i^{\dot\beta}
  +\frac{1}{2}(\sigma^m)^{\dot\alpha}_{\ \ \dot\beta}[a_{i\alpha\dot\alpha},
  a_i^{\alpha\dot\beta}]-\xi^m\right)^2+\left(
  (\sigma^I)^{A}_{\ \ B}\Phi_{iA}\bar\Phi_i^B-
  (\sigma^I)^{A}_{\ \ B}\bar\Phi_{i-1}^B\Phi_{i-1,A}\right)^2\nonumber\\
  &&\hspace{1.3cm}\left.+|\Phi_{iA}q_{i\dot\alpha}|^2+
  |\Phi^\dag_{i+1,A}q_{i\dot\alpha}|^2+|\Phi_{iA}a_{i\alpha\dot\beta}-
  a_{i-1,\alpha\dot\beta}\Phi_{iA}|^2\right]
\end{eqnarray}
where $\xi^3=\xi_{\mathbb{R}}$ and $\xi^1+i\xi^2\sim\xi_{\mathbb{C}}$, with manifest
$SU(2)_{R1}\times SU(2)_{R2}$ symmetry.

Note that with nonzero $\xi^m$, $q_{i\dot\alpha}$
fields are required to be nonzero at low energy, 
which lift the twisted Higgs branch of $\Phi_{iA}$.
Namely, even if $n_1=n_2=\cdots=n_N$, they cannot combine and leave the NS5-branes
unlike the $\mathcal{N}=(4,4)$ model. Also, the previous $(4,4)$ Coulomb
branch fields $a_{\alpha\dot\beta}$ form $(0,4)$ Higgs branch fields, together with
new degrees $q_{\dot\alpha}$. The $(0,4)$ setting will thus be computing the correct 
little string elliptic genus. However, the $(0,4)$ elliptic genus will also capture 
a subtle trace of the presence of a D6-brane from the sector with $n_1=n_2=\cdots=n_N$, 
in which case the D2-branes make full windings along $x^{11}$. This can be easily
accounted for and factored out, after which we shall be obtaining the IIA little 
string index. We shall explain this in section 4.

\subsection{The elliptic genus of IIA little strings}
\label{subsec:Index of IIA LST}

We define the index of IIA little string theory wrapping a spatial circle along $x^{5}$, as follows,
\begin{equation}
	Z_{\rm IIA}(\alpha_{i},\epsilon_{\pm},m;q',w') =\Tr\Big[(-1)^{F}q'^{H_{L}}\bar{q}'^{H_{R}}w'^{k}e^{ 2 \pi i \alpha_{i}\Pi_{i}} e^{2 \pi i\epsilon_{-}(2J_{L1})}e^{2 \pi im (2J_{2L})}e^{2 \pi i\epsilon_{+}(2J_{1R}+2J_{2R})}\Big]\;.
\end{equation}
$\Pi_{i}$ are charges of the self-dual tensor fields, supported on each M5-brane, with the chemical potentials, $\alpha_{i}$.  $q'$ are the fugacity variable counting the number of momentum, and $w'$ the winding fugacity of the IIA little strings. $2 \pi \alpha_{i}R_{M}$
are the positions of the 5-branes along $x^{11}$.

An M2-brane suspended between the $i$'th interval between the M5-branes, $(\alpha_{i},\alpha_{i+1})$, carries nonzero charges $\Pi_{i}=1$ and $\Pi_{i+1}=-1$.
The charges $e_i-e_{i+1}$ form the simple roots of the $A_{N-1}$
algebra, for $i=1,\cdots,N-1$. The last charge $e_N-e_1$, is accompanied by an
extra winding, so the whole $N$ roots become the simples roots of $\hat{A}_{N-1}$.
The fugacity variables corresponding to these simple roots are given by
\begin{equation}
 	v_{1}\equiv e^{2 \pi i \alpha_{12}}\;,\;\;v_{2}\equiv e^{2 \pi i \alpha_{23}}\;,\;\;\cdots\;,\;\;v_{N-1}\equiv e^{2 \pi i \alpha_{N-1,N}}\;,\;\;v_{N} \equiv e^{2 \pi i \alpha_{N,N+1}}=e^{2 \pi i \alpha_{N,1}}w'\;.
\end{equation}
where $\alpha_{ij}=\alpha_{i}-\alpha_{j}$. For convenience, we introduce
$\alpha_{N+1}$, where $e^{-2 \pi i\alpha_{N+1}} = e^{-2 \pi i \alpha_{1}}w'$.

For large $R_A$, the low energy degrees living on the winding strings decouple from
the 6d degrees on the 5-branes. So the index of the IIA little string theory on $\mathbb{R}^{4}\times T^{2}$ factorizes as
\begin{equation}
	Z_{\rm IIA}(\alpha_{i},\epsilon_{\pm},m;q',w') = Z^{\rm IIA}_{\rm mom}
(\epsilon_{\pm},m;q')Z^{\rm IIA}_{\rm string}(\alpha_{i},\epsilon_{\pm},m;q',w')\;.
\end{equation}
$Z_{\rm mom}^{\rm IIA}$ comes from the momenta on $N$ separated  M5-branes wrapping a spatial circle. Unlike the IIB perturbative index
$Z^{\rm IIB}_{\rm pert}$ which had massive W-boson contributions, the IIA 5-brane
does not have extra massive particle states in it (because it only has strings).
So this contribution should factorize into $N$ single 5-brane contributions.
It can be computed either from $N$ Abelian tensor multiplet, or equivalently
from the multiple D0-brane index bound to a single D4-brane \cite{Kim:2011mv}.
The result is
\begin{equation}
 	Z^{\rm IIA}_{\rm mom}(\epsilon_{\pm},m;q') = PE\left[N I_{-}(\epsilon_{1,2},m)\frac{q'}{1-q'}\right]\;,
\end{equation}
where
\begin{equation}
	\label{I-}
	I_{-}(\epsilon_{1,2},m) \equiv \frac{\sinh \frac{2\pi i(m + \epsilon_-)}{2}\sinh\frac{2\pi i (m - \epsilon_-)}{2}}{\sinh \frac{2\pi i \epsilon_{1}}{2}\sinh \frac{2\pi i \epsilon_{2}}{2}}\;
\end{equation}
with $\epsilon_\pm=\frac{\epsilon_1\pm\epsilon_2}{2}$.

The contribution $Z^{\rm IIA}_{\rm string}$ comes from the elliptic genera of the $(0,4)$ gauge theory theory that we have explained in the previous subsection.
This elliptic genus can be computed by the contour integral using the Jeffrey-Kirwan residues \cite{Benini:2013nda,Benini:2013xpa}, or the refined topological vertex method with $(p,q)$-fivebrane web obtained by T-dualizing the branes along $x^5$ 
\cite{Haghighat:2013gba}.
By summing up the elliptic genera over all possible $n_{i}$ numbers, one obtains 
$Z_{\rm string}^{\rm IIA}$. The result is labeled by sets of $N$ Young diagrams, 
$\{Y_{1},...,Y_{N}\}$, where $|Y_{i}| = n_{i}$, 
\begin{align}
    Z^{\rm IIA}_{\rm string}(\alpha_{i},\epsilon_{\pm},m;q',w')  &=\sum_{n_{i}=0}^{\infty}e^{2 \pi i \sum_{i=1}^{N}n_{i} \alpha_{i,i+1}}
    Z^{(n_{1},...,n_{N})}_{\rm string}(\epsilon_{\pm},m;q') \nn \\
    &=\sum_{n_{i}=0}^{\infty}(v_{1})^{n_{1}}(v_{2})^{n_{2}}\cdots(v_{N})^{n_{N}}Z^{(n_{1},...,n_{N})}_{\rm string}(\epsilon_{\pm},m;q').
\end{align}
$Z^{(n_{1},...,n_{N})}_{\rm string}(\epsilon_{\pm},m;q')$ is the elliptic genus 
with fixed $n_{i}$'s, which is given by
\begin{align}
	\label{IIAstring}
    Z^{(n_{1},...,n_{N})}_{\rm string}(\epsilon_{\pm},m;q')   &=\sum_{\{Y_{1},\cdots,Y_{N}\};|Y_{i}|=n_{i}}\prod_{i=1}^{N}\prod_{(a,b)\in Y_{i}}\frac{\theta_{1}(q';E_{i,i+1}^{(a,b)}-m+\epsilon_{-})\theta_{1}(q';E_{i,i-1}^{(a,b)}+m+\epsilon_{-})}{\theta_{1}(q';E_{i,i}^{(a,b)}+ \epsilon_{1})\theta_{1}(q';E_{i,i}^{(a,b)}-\epsilon_{2})}\;,
\end{align}
where
\begin{equation}
    E_{ij}^{(a,b)}=(Y_{i,a}-b)\epsilon_{1}-(Y^{T}_{j,b}-a)\epsilon_{2}\;,\;\;E^{(a,b)}_{i, N+1} =E^{(a,b)}_{i, 1} \;.
\end{equation}
$(a,b)$ denotes the position of each box in a Young diagram. $Y_{a,i}$ is the length of 
the $a$'th row of the Young diagram $Y_{i}$. $Y_{a,i}^{T}$ is the length of the $a$'th 
column of $Y_{i}$.

As we emphasized earlier in this section, the contribution from the $(4,4)$ Higgs branch
(or the $(0,4)$ twisted Higgs branch) formed by $\Phi_A^{(i)}$ is not completely
decoupled. We shall explain at the beginning of section 4 what contribution we expect 
to get from this decoupled sector.

\section{T-duality of protected little string spectra}

\begin{figure}[t!]
	\begin{center}
    \includegraphics[width=17cm]{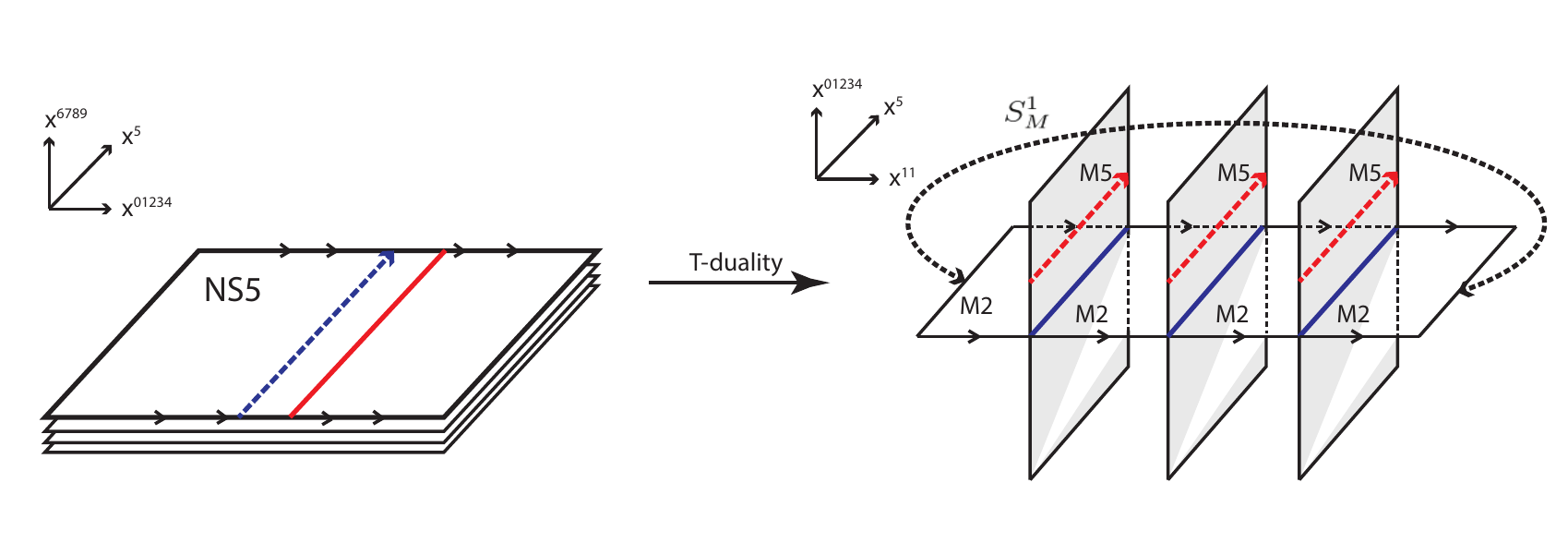}
	\caption{T-duality and M-theory uplift of the IIB setting.
The solid lines represent the winding little strings. The dashed lines represent
the momentum along the circle.}\label{T-duality}
	\end{center}
\end{figure}

The IIB little string theory on a circle is supposed to be T-dual to IIA little 
string theory on the dual circle, with the radiii related by 
$R_{\rm A} =\frac{\alpha'}{R_{\rm B}}$. The winding IIB little strings 
on $S^{1}_{\rm B}$ is dual to the momentum on $S^{1}_{\rm A}$, and
vice versa. Their BPS masses agree with each other, since
\begin{equation}
	m_{\rm IIB\;winding} = \frac{2 \pi R_{\rm B}}{2 \pi \alpha'} = \frac{R_{\rm B}}{\alpha'}=\frac{1}{R_{\rm A}} = m_{\rm IIA\;momentum}\;.
\end{equation}
The fractional momenta of IIB little string theory are dual to the fractional winding 
numbers of IIA little strings,
\begin{align}
	m_{\rm IIB\;KK} =\frac{\alpha_{i,i+1}}{R_{\rm B}}\xrightarrow{\rm T-dual}\left(\alpha_{i,i+1}\right)\frac{R_{\rm A}}{\alpha'}&= \alpha_{i,i+1}(2 \pi R_{\rm A})T_{\rm F1}\ , \nn 
\end{align}
where $\alpha_{ij}=\alpha_{i}-\alpha_{j}$. T-duality between two little string theories is demonstrated by Fig. \ref{T-duality}.

T-duality between IIA and IIB little string theories would naively imply
\begin{equation}
  Z_{\rm IIA}(\alpha_{i},\epsilon_{\pm},m;q',w')|_{q'\rightarrow w, w' \rightarrow q}
  =Z_{\rm IIB}(\alpha_{i},\epsilon_{\pm},m;q,w)\;.
\end{equation}
As we stated at the end of section 3.1, using an alternative 2d $(0,4)$ gauge
theory to compute the IIA elliptic genus (and thus the IIA little string index) will
leave a subtle trace of the fact that we made a UV deformation of the gauge theory
by putting an extra D6-brane. A spectrum change will happen in a sector with full
wound D2-branes along $x^{11}$, namely with states carrying the factors of
fugacities $w^\prime$ but not $\alpha_i$'s. Let us first explain this small subtlety 
in our IIA calculation.

On the IIB side, consider the single particle states with zero electric charges 
(no dependence on $\alpha_i$) and zero winding ($w^0$ order). This contribution 
is contained in
the $Z_{\rm pert}^{\rm IIB}$ factor. In particular, it comes from the $N$ 
Cartan modes of the 6d $U(N)$ SYM. Their partition function is given by
\begin{equation}\label{IIB-neutral}
  PE\left[NI_+(\epsilon_{1,2},m)\frac{q}{1-q}\right]\ .
\end{equation}
In the IIA side, this will correspond to a sector with fully wound D2-branes
along $x^{11}$, at $(q^\prime)^0$ order with $n_1=n_2=\cdots=n_N$. At $\xi^m=0$ 
in section 3.1, there
is an extra twisted Higgs branch which meets the Higgs branch of our interest.
The former sector will represent the little strings leaving NS5-branes.
The two sectors would decouple in IR, but
the 2d gauge theory contains both in its Hilbert space. Now by turning on the
FI term $\xi^m$, the continuum of twisted Higgs branch will be lifted. However,
after turning on $\xi^m$, it often happens that there appear extra bound states of
the continuum degrees
with the remaining 2d strings of our interest. For instance, see \cite{Hwang:2014uwa}
and references therein for many occasions in which extra bound states occur at nonzero
FI parameters. From the $(0,4)$ computation of the IIA side, we shall find
\begin{equation}\label{IIA-neutral}
  PE\left[I_-\frac{w^\prime}{1-w^\prime}+(N-1)I_+\frac{w^\prime}{1-w^\prime}\right]
  =PE\left[NI_+(\epsilon_{1,2},m)\frac{w^\prime}{1-w^\prime}\right]\cdot
  Z_{\rm extra}(w^\prime)
\end{equation}
in the same sector, instead of (\ref{IIB-neutral}),
where $Z_{\rm extra}(w^\prime)\equiv\prod_{n=1}^\infty\frac{1}{1-(w^\prime)^n}
\sim\eta(w^\prime)^{-1}$. We shall give an account for why $Z_{\rm extra}$ should
be appearing due to our $(0,4)$ deformation of the UV theory. With this understood,
we should define the true IIA index as the expression computed in section 3.2 divided
by $Z_{\rm extra}$. We call this
\begin{equation}
  \hat{Z}_{\rm IIA}(\alpha_i,\epsilon_{1,2},m,q^\prime,w^\prime)=
  \frac{{Z}_{\rm IIA}(\alpha_i,\epsilon_{1,2},m,q^\prime,w^\prime)}{Z_{\rm extra}(w^\prime)}.
\end{equation}
We shall find that
\begin{equation}\label{T-duality-final}
  \hat{Z}_{\rm IIA}(\alpha_{i},\epsilon_{\pm},m;q',w')
  |_{q'\rightarrow w, w' \rightarrow q} = Z_{\rm IIB}(\alpha_{i},\epsilon_{\pm},m;q,w)\;,
\end{equation}
which we checked for the cases with $N=1,2,3$. This will establish the T-duality
of the strong-coupling little string spectra via the elliptic genus calculus.

We first explain (or at least heuristically understand) how
$Z_{\rm extra}$ would be appearing in our $(0,4)$ calculus.
Consider the sector with $n_1=\cdots=n_N\equiv n$, forming $n$ full winding
branes which have the right quantum number to leave the NS5-branes. We weight
$n$ windings by $(w^\prime)^n$, and relax the constraint on fixed $n$. We would like
to count the BPS bounds of these strings with $N$ NS5-branes directly, not using the
elliptic genus formula of
\cite{Benini:2013nda,Benini:2013xpa}. We shall do so without and with one D6-brane,
to clearly compare. For convenience, we T-dualize along the $x^{11}$ circle, and obtain
many D1-branes along $05$, $N$-centered Taub-NUT on $678$ and $11$ circle, and optionally
a D5-brane along $012345$ with $B_{\mu\nu}$ (FI term) on $1234$. Firstly, without D5,
any number $n$ of wrapped D1-branes can form a bound state of
multiply wound single string. For each massive particle of this sort, we study
its ground state wavefunction on the $N$-centered Taub-NUT.
This space has $N$ normalizable harmonic forms, so that there could be
$N$ possible bound states of the original $N$ NS5-branes with this particle.
The index for this particle is thus $NI_+(\epsilon_{1,2},m)(w^\prime)^{n}$,
where $N$ comes from $N$ normalizable harmonic forms. The $I_+$ factor appears
because this is exactly the same type of bound states as the half-BPS W-bosons
in SYM, as in the IIB setting. Summing over $n$ and
considering the multi-particle Hilbert space, one exactly obtains (\ref{IIB-neutral})
with $q$ replaced by $w^\prime$. Note that we have arrived at this conclusion
by a direct counting, without any deformation by continuous parameters, so this
should be part of the IIA little string index.

Now we consider the same problem after placing
one D5-brane with FI parameter ($B_{\mu\nu}$ background). The setting of section 3
was that D6 and $N$ NS5-branes are placed at the same point of $\mathbb{R}^3$
in the $678$ directions. Now T-dualizing along $x^{11}$, one finds a D5-brane on
top of the $\mathbb{R}^4/\mathbb{Z}_N$ singularity of unresolved Taub-NUT.
Now, among the $N$ normalizable harmonic forms of $N$-centered Taub-NUT, $N-1$
of them are supported at the $Z_N$ singularity, where D5 is sitting. Since the
fully winding D1-branes are forced to be bound to D5 at the tip due to the FI
parameter, D1-branes stuck to D5 can still assume
one of these $N-1$ bound state wavefunctions. The multi-particle index of the
bounds is $PE[(N-1)I_+\frac{w^\prime}{1-w^\prime}]$. However, the last
normalizable harmonic form of Taub-NUT is not localized at the tip, so D1-branes
confined to D5 cannot be in this bound state. (The forbidden wavefunction is 
in the twisted Higgs branch.) This accounts for
the second term on the left hand side of (\ref{IIA-neutral}). Now, note that
$n$ D1-branes can also form threshold bounds with single D5-brane, whose partition
function is given by $PE[I_-\frac{w^\prime}{1-w^\prime}]$ \cite{Kim:2011mv}. 
(This extra contribution is also from the twisted Higgs branch, since 
the D2-D6 bounds still exist after displacing D6-NS5's.) This
explains the first term of (\ref{IIA-neutral}), and thus the origin of 
$Z_{\rm extra}$. By the discussions of this paragraph, it clearly comes from
having D6-brane and nonzero FI parameter, causing extra bound states or 
destroying some in the twisted Higgs branch.
So with this understood, T-duality would imply (\ref{T-duality-final}).

\subsection{One NS5-brane}

We start by considering the index of the $U(1)$ IIB theory,
although this should be a free theory. The perturbative contribution 
is given by
\begin{equation}
	Z^{\rm IIB}_{\mathrm{pert}}(\epsilon_{\pm},m;q) = PE\left[I_{+}(\epsilon_{\pm},m)\frac{q}{1-q} \right].
\end{equation}
The $U(1)$ instanton string partition function is given by
\begin{equation}
	Z^{\rm IIB}_{\mathrm{string}}(\epsilon_{\pm},m;q,w) = \sum_{k=0}w^{k}Z_{k}(\epsilon_{\pm},m;q)\ ,
\end{equation}
where
\begin{equation}
 	Z_{k} = \sum_{Y :|Y|=k}\prod_{s \in Y}\frac{\theta_{1}\left(q;E(s)+m- \epsilon_{-}\right)\theta_{1}\left(q;E(s) - m - \epsilon_{-}\right)}{\theta_{1}\left(q;E(s)-\epsilon_{1}\right)\theta_{1}\left(q;E(s) + \epsilon_{2}\right)}\;,
\end{equation}
with
\begin{equation}
	E(s) = -\epsilon_{1}h(s)+\epsilon_{2}v(s)\;.
\end{equation}
The full index of the $U(1)$ theory is given by
\begin{equation}
	Z_{{\rm IIB}}(\epsilon_{\pm},m;q, w) = Z^{\rm IIB}_{\rm \;pert}(\epsilon_{\pm},m;q)
Z^{\rm IIB}_{\rm \;inst}(\epsilon_{\pm},m;q, w)
\end{equation}
To further explain this index, consider the single instanton string index given by
\begin{equation}
	Z_{1}(\epsilon_{\pm},m;q)=\frac{\theta_{1}\left(q;m \pm \epsilon_{-}\right)}{\theta_{1}\left(q;\epsilon_{1}\right)\theta_{1}\left(q;\epsilon_{2}\right)}\;.
\end{equation}
where $\theta_{1}(q;a\pm b) \equiv \theta_{1}(q;a+b)\theta_{1}(q;a-b)$.
In terms of $Z_1$, we find that the multi-instanton string index is given by
the Hecke transformation of $Z_{1}$,
\begin{equation}\label{hecke}
 	Z_{\rm inst}(\epsilon_{\pm},m;q,w) = \exp\left[\sum_{n=1}^{\infty}\frac{1}{n}w^{n}\sum_{\overset{ad=n}{\underset{a,d \in \mathbb{Z}}{}}}\sum_{b({\rm mod}\;d)}Z_{1}\left(a \epsilon_{\pm},a m;\frac{a \tau + b}{d}\right)\right]
\end{equation}
where $q=e^{2 \pi i \tau}$. This is checked up to high orders in $w$ and $q$.

The partition function given by the Hecke transformation appears,
for instance, in conformal field theories on symmetric product
target spaces. This is closely related to the fact that the moduli-space of $U(1)$
multi-instantons is a symmetric product of the
single instanton moduli space $\mathbb{R}^{4}$. More precisely,
the symmetric product CFT was suggested to be the theory at nonzero world-sheet
theta angle $\theta = \pi$ \cite{Witten:1997yu,Seiberg:1999xz}. Since the elliptic
genus would be insensitive to the continuous parameters, away from $\zeta^I=0$,
$\theta=0$, it is natural to have (\ref{hecke}).

\begin{figure}[t!]
	\begin{center}
    \includegraphics[width=14cm]{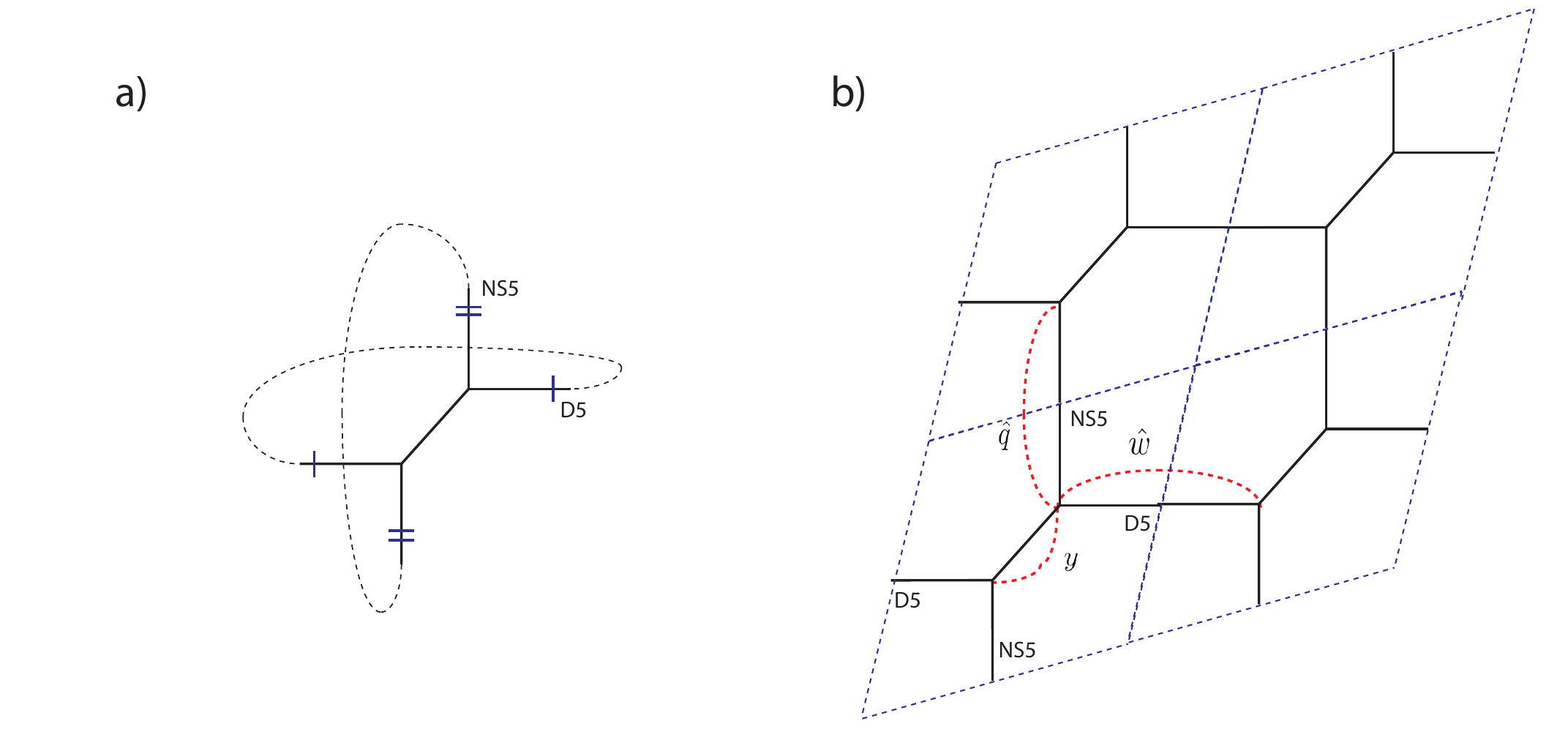}
	\caption{a) $(p,q)$ fivebranes web dual to rank 1 little string theory. b) Triality between three K\"ahler parameters, $\hat{q}=qy^{-1}$, $\hat{w}=wy^{-1}$, and $y$.}\label{Fig:Rank1-5branes}
	\end{center}
\end{figure}

On the IIA side, the 2d $\mathcal{N}=(0,4)$
quiver gauge theory itself has an enhanced $\mathcal{N}=(4,4)$ SUSY, and becomes
precisely the same to the 2d $\mathcal{N}=(4,4)$ ADHM gauge theory for IIB
strings. Therefore, 
\begin{align}
	Z^{\rm IIA}_{\rm string}(\epsilon_{\pm},m;q^\prime,w^\prime)=
Z^{\rm IIB}_{\rm inst}(\epsilon_\pm,m,q^\prime,w^\prime)\ .
\end{align}
The extra factor $Z^{\rm IIA}_{\rm mom}$ on the IIA side is given by
\begin{equation}
	Z^{\rm IIA}_{\rm mom}(\epsilon_{\pm},m;q') = PE\left[I_{-}(\epsilon_{\pm},m)\frac{q'}{1-q'} \right]=
Z^{\rm IIB}_{\rm pert}(\epsilon_\pm,m,q^\prime)
Z_{\rm extra}(q^\prime)\ .
\end{equation}
So the T-duality relation (\ref{T-duality-final}) is equivalent to 
$Z_{\rm IIA}(\epsilon_{\pm},m;q',w')$ being invariant under the exchange of 
$q'$ and $w'$. The last property is in fact true,
which can be understood as the geometric duality of the 5-brane web obtained 
by T-dualizing our IIA brane setting along $x^5$ \cite{Hollowood:2003cv}, 
as shown in Fig. \ref{Fig:Rank1-5branes}.
If we write $Z_{\rm IIA}(\epsilon_{\pm},m;q',w')$ as
\begin{equation}
	Z_{\rm IIA}(\epsilon_{\pm},m;q',w') = PE\Big[I_{-}(\epsilon_{\pm},m)z_{\rm sp}(\epsilon_{\pm},m,q^\prime,w^\prime)\Big]\;,
\end{equation}
$z_{\rm sp}(\epsilon_{\pm},m)$ is given by
\begin{align}
	&z_{\rm sp}(\epsilon_{\pm},m;q',w') = (q'+w')+(q'^{2}+w'^{2})+(q'w')\left[t u+\frac{t}{u}+\frac{1}{t u}+\frac{u}{t}-u y-\frac{y}{u}-\frac{u}{y}-\frac{1}{u y}\right] \nn \\
	&+q'^{3}+w'^{3}+(q'^{2}w'+q'w'^{2})\left[t^2 u^2+\frac{t^2}{u^2}+\frac{u^2}{t^2}+\frac{1}{t^2 u^2}+t^2+\frac{1}{t^2}-t u^2 y-\frac{t y}{u^2}-\frac{t u^2}{y}-\frac{t}{u^2 y} \right.\nn \\
	&-\frac{y}{t u^2}-\frac{u^2}{t y}-\frac{1}{t u^2 y}-\frac{u^2 y}{t}+t u+\frac{t}{u}+\frac{1}{t u}+\frac{u}{t}-2 t y-\frac{2 t}{y}-\frac{2}{t y}-\frac{2 y}{t}+2 u^2+\frac{2}{u^2}-u y-\frac{y}{u} \nn \\
	&\left.-\frac{u}{y}-\frac{1}{u y}+y^2+\frac{1}{y^2}+4\right]+(q'^{4}+w'^{4})+(q'^{3}w'+q'w'^{3})\left[t^3 u^3+\frac{t^3}{u^3}+\frac{u^3}{t^3}+\frac{1}{t^3 u^3}+t^3 u+\frac{t^3}{u} \right. \nn \\
	&+\frac{u}{t^3}+\frac{1}{t^3 u}-t^2 u^3 y-\frac{t^2 y}{u^3}-\frac{t^2 u^3}{y}-\frac{t^2}{u^3 y}-\frac{u^3 y}{t^2}-\frac{y}{t^2 u^3}-\frac{u^3}{t^2 y}-\frac{1}{t^2 u^3 y}+t^2 u^2+\frac{t^2}{u^2}+\frac{u^2}{t^2} \nn \\
	&+\frac{1}{t^2 u^2}-2 t^2 u y-\frac{2 t^2 y}{u}-\frac{2 t^2 u}{y}-\frac{2 t^2}{u y}-\frac{2 u y}{t^2}-\frac{2 y}{t^2 u}-\frac{2 u}{t^2 y}-\frac{2}{t^2 u y}+2 t^2+\frac{2}{t^2}+2 t u^3+\frac{2 t}{u^3} \nn
\end{align}
\begin{align}
	&+\frac{2}{t u^3}+\frac{2 u^3}{t}-2 t u^2 y-\frac{2 t y}{u^2}-\frac{2 t u^2}{y}-\frac{2 t}{u^2 y}-\frac{2 y}{t u^2}-\frac{2 u^2}{t y}-\frac{2}{t u^2 y}-\frac{2 u^2 y}{t} +t u y^2+\frac{t y^2}{u}+\frac{t u}{y^2}\nn \\
	&+\frac{t}{u y^2}+\frac{y^2}{t u}+\frac{u}{t y^2}+\frac{1}{t u y^2}+\frac{u y^2}{t}+6 t u+\frac{6 t}{u}+\frac{6}{t u}+\frac{6 u}{t}-4 t y-\frac{4 t}{y}-\frac{4}{t y}-\frac{4 y}{t}-u^3 y-\frac{y}{u^3} \nn \\
	&\left.-\frac{u^3}{y}-\frac{1}{u^3 y}+u^2 y^2+\frac{y^2}{u^2}+\frac{u^2}{y^2}+\frac{1}{u^2 y^2}+4 u^2+\frac{4}{u^2}-5 u y-\frac{5 y}{u}-\frac{5 u}{y}-\frac{5}{u y}+2 y^2+\frac{2}{y^2}+8\right] \nn \\
	&+(q'^{2}w'^{2})\left[u^4 t^4+u^2 t^4+\frac{t^4}{u^2}+\frac{t^4}{u^4}+t^4+u^3 t^3+2 u t^3-u^4 y t^3-2 u^2 y t^3-2 y t^3+\frac{2 t^3}{u}-\frac{2 y t^3}{u^2} \right. \nn \\
	&+\frac{t^3}{u^3}-\frac{y t^3}{u^4}-\frac{u^4 t^3}{y}-\frac{2 u^2 t^3}{y}-\frac{2 t^3}{y}-\frac{2 t^3}{u^2 y}-\frac{t^3}{u^4 y}+2 u^4 t^2+7 u^2 t^2+u^2 y^2 t^2+\frac{y^2 t^2}{u^2}+y^2 t^2 \nn \\
	&-2 u^3 y t^2-5 u y t^2-\frac{5 y t^2}{u}+\frac{7 t^2}{u^2}-\frac{2 y t^2}{u^3}+\frac{2 t^2}{u^4}-\frac{2 u^3 t^2}{y}-\frac{5 u t^2}{y}-\frac{5 t^2}{u y}-\frac{2 t^2}{u^3 y}+\frac{u^2 t^2}{y^2}+\frac{t^2}{y^2} \nn \\
	&+\frac{t^2}{u^2 y^2}+9 t^2+5 u^3 t+u^3 y^2 t+4 u y^2 t+\frac{4 y^2 t}{u}+\frac{y^2 t}{u^3}+15 u t-u^4 y t-7 u^2 y t-12 y t+\frac{15 t}{u} \nn \\
	&-\frac{7 y t}{u^2}+\frac{5 t}{u^3}-\frac{y t}{u^4}-\frac{u^4 t}{y}-\frac{7 u^2 t}{y}-\frac{12 t}{y}-\frac{7 t}{u^2 y}-\frac{t}{u^4 y}+\frac{u^3 t}{y^2}+\frac{4 u t}{y^2}+\frac{4 t}{u y^2}+\frac{t}{u^3 y^2} +\frac{2 u^4}{t^2}
	\nn \\
	&+\frac{u^4}{t^4}+2 u^4+\frac{u^3}{t^3}-u y^3+\frac{7 u^2}{t^2}+\frac{u^2}{t^4}+12 u^2+\frac{u^2 y^2}{t^2}+2 u^2 y^2+\frac{y^2}{t^2}+\frac{4 y^2}{t u}+\frac{2 y^2}{u^2}+\frac{y^2}{t^2 u^2}+\frac{y^2}{t u^3} \nn \\
	&+5 y^2+\frac{2 u}{t^3}-4 u^3 y-14 u y-\frac{2 u^3 y}{t^2}-\frac{5 u y}{t^2}+\frac{9}{t^2}-\frac{u^4 y}{t^3}-\frac{2 u^2 y}{t^3}-\frac{2 y}{t^3}+\frac{1}{t^4}-\frac{y^3}{u}-\frac{14 y}{u}+\frac{15}{t u} \nn \\
	&-\frac{5 y}{t^2 u}+\frac{2}{t^3 u}+\frac{12}{u^2}-\frac{7 y}{t u^2}+\frac{7}{t^2 u^2}-\frac{2 y}{t^3 u^2}+\frac{1}{t^4 u^2}-\frac{4 y}{u^3}+\frac{5}{t u^3}-\frac{2 y}{t^2 u^3}+\frac{1}{t^3 u^3}+\frac{2}{u^4}-\frac{y}{t u^4}+\frac{2}{t^2 u^4} \nn \\
	&-\frac{y}{t^3 u^4}+\frac{1}{t^4 u^4}-\frac{4 u^3}{y}-\frac{14 u}{y}-\frac{u^4}{t y}-\frac{7 u^2}{t y}-\frac{12}{t y}-\frac{2 u^3}{t^2 y}-\frac{5 u}{t^2 y}-\frac{u^4}{t^3 y}-\frac{2 u^2}{t^3 y}-\frac{2}{t^3 y}-\frac{14}{u y}-\frac{5}{t^2 u y} \nn \\
	&-\frac{7}{t u^2 y}-\frac{2}{t^3 u^2 y}-\frac{4}{u^3 y}-\frac{2}{t^2 u^3 y}-\frac{1}{t u^4 y}-\frac{1}{t^3 u^4 y}+\frac{2 u^2}{y^2}+\frac{5}{y^2}+\frac{u^3}{t y^2}+\frac{4 u}{t y^2}+\frac{u^2}{t^2 y^2}+\frac{1}{t^2 y^2}+\frac{4}{t u y^2} \nn \\
	&+\frac{2}{u^2 y^2}+\frac{1}{t^2 u^2 y^2}+\frac{1}{t u^3 y^2}-\frac{u}{y^3}-\frac{1}{u y^3}+22+\frac{5 u^3}{t}+\frac{u^3 y^2}{t}+\frac{4 u y^2}{t}+\frac{15 u}{t}-\frac{u^4 y}{t}-\frac{7 u^2 y}{t} \nn \\
	&\left.-\frac{12 y}{t} \right]+ \cdots.
\end{align}
where $t= e^{2 \pi i \epsilon_{+}}$, $u=e^{2 \pi i \epsilon_{-}}$, $y = e^{2 \pi i m}$. We checked the symmetry of $q'\leftrightarrow w'$ exchange
up to 5th orders in $q'$ and $w'$.

Furthermore, defining the following variables,
\begin{equation}
	\hat{q} = q y^{-1}\;,\;\;\hat{w} = w y^{-1}\;.
\end{equation}
triality of exchanging $(\hat{q}, \hat{w},y)$ has been discovered in \cite{Hollowood:2003cv}. This is also a geometric duality of Fig. \ref{Fig:Rank1-5branes}. 
Triality is simply realized on the universal covering of the torus, as a subgroup 
of $Sp(4, Z)$ duality. To deal with $(\hat{q}, \hat{w},y)$ in equal footing, we redefine 
the index, including extra perturbative contributions at $y\ll 1$, as
\begin{equation}
	\tilde{Z}(\epsilon_{\pm};\hat{q},\hat{w},y) = PE\left[I_{\rm com}(\epsilon_{\pm})y\right]Z_{\rm IIA}\ .
\end{equation}
$I_{\rm com}(\epsilon_{\pm})$ is given by
\begin{equation}
	\label{Icom}
	I_{\rm com}(\epsilon_{\pm}) = \frac{1}{2 \sinh\frac{2 \pi i \epsilon_{1}}{2}2 \sinh\frac{2 \pi i \epsilon_{2}}{2}} =  \frac{t}{(1-tu)(1-tu^{-1})}.
\end{equation}
Writing $\tilde{Z}$ as
\begin{equation}
	\tilde{Z}(\epsilon_{\pm};\hat{q},\hat{w},y) = PE\Big[I_{\rm com}\tilde{z}_{\rm sp}(\epsilon_{\pm};\hat{q},\hat{w},y)\Big],
\end{equation}
$\tilde{z}_{\rm sp}$ is given by
\begin{align}
	&\hat{z}_{sp}(\epsilon_{\pm};\hat{q},\hat{w},y) = \hat{q}+\hat{w}+y-(u+u^{-1})(\hat{q}\hat{w}+\hat{q}y+\hat{w}y)+\frac{(1 + u^2) (t + u + t^2 u + t u^2)}{t u^2}\hat{q}\hat{w}y\nn \\
	 &+(\hat{q}^{2}\hat{w}+\hat{q}\hat{w}^{2}+\hat{q}^{2}y+\hat{q}y^{2}+\hat{w}^{2}y+\hat{w}y^{2})-(u+u^{-1})(\hat{q}^{2}\hat{w}^{2}+\hat{q}^{2}y^{2}+\hat{w}^{2}y^{2}) \nn \\
	&-\frac{\left(u^2+1\right) \left(t^2 \left(u^2+1\right)+2 t u+u^2+1\right)}{t u^2}\hat{q}\hat{w}y(\hat{q}+\hat{w}+y) \nn \\
	 &+(\hat{q}^{3}\hat{w}^{2}+\hat{q}^{2}\hat{w}^{3}+\hat{q}^{3}y^{2}+\hat{q}^{2}y^{3}+\hat{w}^{3}y^{2}+\hat{w}^{2}y^{3}) +\frac{(1 + u^2) (t + u + t^2 u + t u^2)}{t u^2}\hat{q}\hat{w}y (\hat{q}^2+\hat{w}^2+y^{2}) \nn \\
	&+\frac{t^4 \left(u^5+u^3+u\right)+t^3 \left(u^6+4 u^4+4 u^2+1\right)}{t^2 u^3}\hat{q}\hat{w}y(\hat{q}\hat{w}+\hat{q}y+\hat{w}y) \nn \\
	&+\frac{t^2 \left(3 u^4+7 u^2+3\right) u+t \left(u^6+4 u^4+4 u^2+1\right)+u^5+u^3+u}{t^2 u^3}\hat{q}\hat{w}y(\hat{q}\hat{w}+\hat{q}y+\hat{w}y) \nn \\
	&-(u+u^{-1})(\hat{q}^{3}\hat{w}^{3}+\hat{q}^{3}y^{3}+\hat{w}^{3}y^{3}) \nn \\
	&-\frac{\left(u^2+1\right) \left(t^4 \left(u^4+u^2+1\right)+3 t^3 \left(u^3+u\right)\right)}{t^2 u^3}\hat{q}\hat{w}y(\hat{q}^{2}\hat{w}+\hat{q}\hat{w}^{2}+\hat{q}^{2}y+\hat{q}y^{2}+\hat{w}^{2}y+\hat{w}y^{2}) \nn \\
	&-\frac{\left(u^2+1\right) \left(2 t^2 \left(u^4+3 u^2+1\right)+3 t \left(u^3+u\right)+u^4+u^2+1\right)}{t^2 u^3}\hat{q}\hat{w}y(\hat{q}^{2}\hat{w}+\hat{q}\hat{w}^{2}+({\rm cyclic})) \nn \\
	&+\cdots
\end{align}
reconfirming the expected triality of \cite{Hollowood:2003cv}.
It is curious to note that the triality implies the T-duality of IIA/IIB strings.

\subsection{Two NS5-branes}

The index of $U(2)$ IIB little string theory is given by
\begin{equation}
  Z_{\rm IIB}(\alpha_{i},\epsilon_{\pm},m;q,w) =
  Z^{\rm IIB}_{\rm pert}(\alpha_{i},\epsilon_{\pm},m;q) Z^{\rm IIB}_{\rm inst}(\alpha_{i},\epsilon_{\pm},m;q,w)\ ,
\end{equation}
where 
\begin{equation}
	Z^{\rm IIB}_{\rm pert}(\alpha_{i},\epsilon_{\pm},m;q) =PE\left[
	I_{+}v_{1} +\left(2I_{+}+ I_{+}(v_{1}+v_{1}^{-1})\right)\frac{q}{1-q}\right]
	=PE\left[I_{+}\frac{v_{1}+v_{2}+2v_{1}v_{2}}{1-v_{1}v_{2}}\right],
\end{equation}
with $v_{1} = e^{2 \pi i \alpha_{12}}$, and $v_{2}\equiv qv_{1}^{-1}$. $I_{+}$ is given by eq.(\ref{I+}). $Z^{\rm IIB}_{\rm inst}$ is given by
\begin{equation}
	Z^{\rm IIB}_{\rm inst}(\alpha_{i},\epsilon_{\pm},m;q,w) =\sum_{k=0}^{\infty}w^{k}Z_{k}(\alpha_{i},\epsilon_{\pm},m;q)\;.
\end{equation}
$Z_{k}$ is obtained from eq.(\ref{Instanton}). 
Expanding $Z^{\rm IIB}_{\rm inst}(\alpha_{i},\epsilon_{\pm},m;q,w)$ with $w$, $v_{1}$, $v_{2} = q v_{1}^{-1}$, one obtains
\begin{align}
	&Z^{\rm IIB}_{\rm inst}(\epsilon_{\pm},m;w,v_{i}) \nn \\
	&= 1-w\frac{2 t (u-y) (u y-1)}{y (t-u) (t u-1)}+w(v_{1}+v_{2})\frac{\left(t^2+1\right) (t-y) (t y-1) (y-u) (u y-1)}{t y^2 (t-u) (t u-1)} \nn \\
	&+w(v_{1}^2+v_{2}^{2})\frac{\left(t^2+1\right) \left(t^4+1\right) (t-y) (t y-1) (y-u) (u y-1)}{t^3 y^2 (t-u) (t u-1)} \nn \\
	&+w v_{1}v_{2} \frac{2 (t-y) (t y-1) (y-u) (u y-1) \left(t^2 u y+t (u-y) (u y-1)+u y\right)}{t u y^3 (t-u) (t u-1)} \nn \\
	&+w (v_{1}^{2}v_{2}+v_{1}v_{2}^{2})\frac{\left(t^2+1\right) (t-y) (t y-1) (y-u) (u y-1)}{t^3 u y^3 (t-u) (t u-1)} \nn \\
	&\times\Big\{-t (t + u) (1 + t u) (1 + y^2) + (t + u + t^2 u) (1 + t (t + u)) y\Big\} +\cdots
\end{align}

The index for the rank 2  IIA little string theory is given by
\begin{equation}
	\hat{Z}_{\rm IIA}= Z_{\rm extra}(q)^{-1} Z^{\rm IIA}_{\rm mom}
(\epsilon_{\pm},m;w)Z^{\rm IIA}_{\rm string}(\alpha_{i},\epsilon_{\pm},m;w,q)\;
\end{equation}
where we inserted $q^\prime=w$, $w^\prime=q$. $Z_{\rm extra}(q)$ is given by
\begin{equation}
	Z_{\rm extra}(q) = PE\left[\frac{q}{1-q}\right] = PE\left[\frac{v_{1}v_{2}}{1-v_{1}v_{2}}\right]\;.
\end{equation}
$Z^{\rm IIA}_{\rm mom}(\epsilon_{\pm},m;w)$ is given by
\begin{equation}
	Z^{\rm IIA}_{\rm N=2\;mom}(\epsilon_{\pm},m;w)=PE\left[2I_{-}(\epsilon_{\pm},m)\frac{w}{1-w} \right]\;.
\end{equation}
$Z^{\rm IIA}_{\rm string}(\alpha_{i},\epsilon_{\pm},m;w,q)$ takes the form of
\begin{equation}
    Z^{\rm IIA}_{\rm string}(\alpha_{i},\epsilon_{\pm},m;w,q)  
    =\sum_{n_{1},n_{2}=0}^{\infty}(v_{1})^{n_{1}}(v_{2})^{n_{2}}Z^{(n_{1},n_{2})}_{\rm string}(\epsilon_{\pm},m;w)\;,\;\;
\end{equation}
Note that $Z^{(n_{1},n_{2})}_{\rm string}(\epsilon_{\pm},m;w) = Z^{(n_{2},n_{1})}_{\rm string}(\epsilon_{\pm},m;w)$, from the symmetry of the quiver. $Z^{(n_{1},n_{2})}_{\rm string}(\epsilon_{\pm},m;w)$ can be easily obtained from (\ref{IIAstring}). 
For instance,
\begin{equation}
	Z^{(1,0)}_{\rm string}(\epsilon_{\pm},m;w)=\frac{\theta_{1}(w,m \pm \epsilon_{+})}{\theta_{1}(w,\epsilon_{1})\theta_{1}(w,\epsilon_{2})}\;,\;\;Z^{(1,1)}_{\rm string}(\epsilon_{\pm},m;w)=\frac{\theta_{1}(w,m\pm \epsilon_{-})^{2}}{\theta_{1}(w,\epsilon_{1})^{2}\theta_{1}(w,\epsilon_{2})^{2}}
\end{equation}
\begin{align}
	&Z^{(2,0)}_{\rm string}(\epsilon_{\pm},m;w)=
	Z^{(1,0)}_{\rm string} \cdot \left(\frac{ \theta _1\left(w;\epsilon _++\epsilon _1\pm m\right)}{\theta _1\left(w;2 \epsilon _1\right) \theta _1\left(w;\epsilon _1-\epsilon _2\right)}-(\epsilon_{1}\leftrightarrow \epsilon_{2})\right)
\end{align}
We write the indices of the IIA/IIB little string theories as
\begin{equation}
	Z_{\rm IIB}(\alpha_{i},\epsilon_{\pm},m;w,v_{i}) = PE\left[I_{\rm com}(t,u)\sum_{i,j,k=0}^{\infty}F^{\rm IIB}_{ijk}(t,u,y)w^{i}v_{1}^{j}v_{2}^{k}\right]\;,
\end{equation}
\begin{equation}
	\hat{Z}_{\rm IIA}(\alpha_{i},\epsilon_{\pm},m;w,v_{i})=PE\left[I_{\rm com}(t,u)\sum_{i,j,k=0}^{\infty}F^{\rm IIA}_{ijk}(t,u,y)w^{i}v_{1}^{j}v_{2}^{k}\right]\;,
\end{equation}
where $I_{\rm com}$ is given by eq.(\ref{Icom}). The coefficients $F^{\rm IIB}_{ijk}(t,u,y)$ are polynomials of $t= e^{2 \pi i \epsilon_{+}}$, $u=e^{ 2 \pi i \epsilon_{-}}$, and $y=e^{ 2 \pi i m}$.
It is easily checked that $F^{\rm IIB}_{ijk}(t,u,y)=F^{\rm IIB}_{ikj}(t,u,y)$.

T-duality implies that $F^{\rm IIA}_{ijk}=F^{\rm IIB}_{ijk}\equiv F_{ijk}$. 
We check T-daulity by comparing  $F^{\rm IIA}_{ijk}$ and $F^{\rm IIB}_{ijk}$. 
We checked the agreements for 
\begin{equation}
	F_{000} =1\;,\;\;F_{010}=-t-\frac{1}{t}+y+\frac{1}{y}\;,\;\;F_{011}=-2 t-\frac{2}{t}+2 y+\frac{2}{y}
\end{equation}
\begin{equation}
	F_{020}=0\;,\;\;F_{021}=-t-\frac{1}{t}+y+\frac{1}{y}\;,\;\;F_{022}=-2 t-\frac{2}{t}+2 y+\frac{2}{y}
\end{equation}
\begin{equation}
	F_{100}=-2 u-\frac{2}{u}+2 y+\frac{2}{y}\;,\;\;
\end{equation}
\begin{align}
	&F_{110}=-t^2u-\frac{t^2}{u}-\frac{u}{t^2}-\frac{1}{t^2 u}+t^2 y+\frac{t^2}{y}+\frac{y}{t^2}+\frac{1}{t^2 y}+t u y +\frac{t y}{u}+\frac{t u}{y}+\frac{t}{u y}+\frac{y}{t u}+\frac{u}{t y}\nn \\
	&+\frac{1}{t u y}+\frac{u y}{t}-t y^2-\frac{t}{y^2}-\frac{1}{t y^2}-\frac{y^2}{t}-2 t-\frac{2}{t}-2 u-\frac{2}{u}+2 y+\frac{2}{y}
\end{align}
\begin{align}
    &F_{111}=-2 t^2 u-\frac{2 u}{t^2}-\frac{2 t^2}{u}-\frac{2}{t^2 u}+2 t^2 y+\frac{2 y}{t^2}+\frac{2}{t^2 y}+\frac{2 t^2}{y}-2 t u^2-\frac{2 u^2}{t}-\frac{2 t}{u^2}-\frac{2}{t u^2}+6 t u y \nn \\
    &+\frac{6 u y}{t}+\frac{6 t y}{u}+\frac{6 y}{t u}+\frac{6 u}{t y}+\frac{6 t}{u y}+\frac{6}{t u y}+\frac{6 t u}{y}-4 t y^2-\frac{4 y^2}{t}-\frac{4 t}{y^2}-\frac{4}{t y^2}-12 t-\frac{12}{t}+2 u^2 y \nn \\
    &+\frac{2 y}{u^2}+\frac{2}{u^2 y}+\frac{2 u^2}{y}-4 u y^2-\frac{4 y^2}{u}-\frac{4 u}{y^2}-\frac{4}{u y^2}-12 u-\frac{12}{u}+2 y^3+\frac{2}{y^3}+14 y+\frac{14}{y}
\end{align}
\begin{align}
    &F_{120}=-t^4 u-\frac{t^4}{u}-\frac{u}{t^4}-\frac{1}{t^4 u}+t^4 y+\frac{t^4}{y}+\frac{y}{t^4}+\frac{1}{t^4 y}+t^3 u y+\frac{t^3 y}{u}+\frac{t^3 u}{y}+\frac{t^3}{u y}+\frac{u y}{t^3}+\frac{y}{t^3 u} \nn \\
    &+\frac{u}{t^3 y}+\frac{1}{t^3 u y}-t^3 y^2-\frac{t^3}{y^2}-\frac{y^2}{t^3}-\frac{1}{t^3 y^2}-2 t^3-\frac{2}{t^3}-2 t^2 u-\frac{2 t^2}{u}-\frac{2 u}{t^2}-\frac{2}{t^2 u}+2 t^2 y+\frac{2 t^2}{y} \nn \\
    &+\frac{2 y}{t^2}+\frac{2}{t^2 y}+t u y+\frac{t y}{u}+\frac{t u}{y}+\frac{t}{u y}+\frac{y}{t u}+\frac{u}{t y}+\frac{1}{t u y}+\frac{u y}{t}-t y^2-\frac{t}{y^2}-\frac{1}{t y^2} -\frac{y^2}{t}-2 t-\frac{2}{t}\nn \\
    &-2 u-\frac{2}{u}+2 y+\frac{2}{y}
\end{align}
\begin{align}
	&F_{121}= - t^4 u +y t^4-\frac{t^4}{u}+\frac{t^4}{y}-u^2 t^3-2 y^2 t^3+3 u y t^3+\frac{3 y t^3}{u}-\frac{t^3}{u^2}+\frac{3 u t^3}{y}+\frac{3 t^3}{u y}-\frac{2 t^3}{y^2} -6 t^3\nn \\
	&+y^3 t^2-3 u y^2 t^2-11 u t^2+2 u^2 y t^2+\frac{2 y t^2}{u^2}+12 y t^2-\frac{3 y^2 t^2}{u}-\frac{11 t^2}{u}+\frac{2 u^2 t^2}{y}+\frac{12 t^2}{y}+\frac{2 t^2}{u^2 y} \nn \\
	&-\frac{3 u t^2}{y^2}-\frac{3 t^2}{u y^2}+\frac{t^2}{y^3}+u y^3 t+\frac{y^3 t}{u}-5 u^2 t-u^2 y^2 t-9 y^2 t+13 u y t+\frac{13 y t}{u}-\frac{y^2 t}{u^2}-\frac{5 t}{u^2}+\frac{13 u t}{y}\nn \\
	&+\frac{13 t}{u y}-\frac{u^2 t}{y^2} -\frac{9 t}{y^2}-\frac{t}{u^2 y^2}+\frac{u t}{y^3}+\frac{t}{u y^3}-24 t+\frac{y^3}{t^2}+\frac{y^3}{t u}+2 y^3-6 u y^2-20 u+\frac{2 u^2 y}{t^2}+4 u^2 y\nn \\
	&+\frac{3 u y}{t^3} +\frac{12 y}{t^2}+\frac{y}{t^4}+\frac{13 y}{t u}+\frac{3 y}{t^3 u}+\frac{4 y}{u^2}+\frac{2 y}{t^2 u^2}+22 y-\frac{3 u y^2}{t^2}-\frac{11 u}{t^2}-\frac{u^2}{t^3}-\frac{2 y^2}{t^3}-\frac{6}{t^3}-\frac{u}{t^4} \nn \\
	&-\frac{6 y^2}{u}-\frac{20}{u}-\frac{3 y^2}{t^2 u}-\frac{11}{t^2 u}-\frac{1}{t^4 u}-\frac{y^2}{t u^2}-\frac{5}{t u^2}-\frac{1}{t^3 u^2}+\frac{4 u^2}{y}+\frac{22}{y}+\frac{13 u}{t y}+\frac{2 u^2}{t^2 y}+\frac{12}{t^2 y}+\frac{3 u}{t^3 y} \nn \\
	&+\frac{1}{t^4 y}+\frac{13}{t u y}+\frac{3}{t^3 u y}+\frac{4}{u^2 y}+\frac{2}{t^2 u^2 y}-\frac{6 u}{y^2}-\frac{u^2}{t y^2}-\frac{9}{t y^2}-\frac{3 u}{t^2 y^2}-\frac{2}{t^3 y^2}-\frac{6}{u y^2}-\frac{3}{t^2 u y^2}-\frac{1}{t u^2 y^2} \nn \\
	&+\frac{2}{y^3}+\frac{u}{t y^3}+\frac{1}{t^2 y^3}+\frac{1}{t u y^3}+\frac{u y^3}{t}-\frac{5 u^2}{t}-\frac{u^2 y^2}{t}-\frac{9 y^2}{t}+\frac{13 u y}{t}-\frac{24}{t}
\end{align}
\begin{align}
    &F_{122}=-2 u t^4+2 y t^4-\frac{2 t^4}{u}+\frac{2 t^4}{y}-2 u^2 t^3-4 y^2 t^3+6 u y t^3+\frac{6 y t^3}{u}-\frac{2 t^3}{u^2}+\frac{6 u t^3}{y}+\frac{6 t^3}{u y}-\frac{4 t^3}{y^2} \nn \\
    &-12 t^3-2 u^3 t^2+2 y^3 t^2-8 u y^2 t^2-32 u t^2+8 u^2 y t^2+\frac{8 y t^2}{u^2}+32 y t^2-\frac{8 y^2 t^2}{u}-\frac{32 t^2}{u}-\frac{2 t^2}{u^3}+\frac{8 u^2 t^2}{y} \nn \\
    &+\frac{32 t^2}{y}+\frac{8 t^2}{u^2 y}-\frac{8 u t^2}{y^2}-\frac{8 t^2}{u y^2}+\frac{2 t^2}{y^3}+6 u y^3 t+\frac{6 y^3 t}{u}-24 u^2 t-8 u^2 y^2 t-36 y^2 t+2 u^3 y t +52 u y t\nn \\
    &+\frac{52 y t}{u}+\frac{2 y t}{u^3}-\frac{8 y^2 t}{u^2}-\frac{24 t}{u^2}+\frac{2 u^3 t}{y}+\frac{52 u t}{y}+\frac{52 t}{u y}+\frac{2 t}{u^3 y}-\frac{8 u^2 t}{y^2}-\frac{36 t}{y^2}-\frac{8 t}{u^2 y^2}+\frac{6 u t}{y^3}+\frac{6 t}{u y^3} \nn \\
    &-88 t-2 u y^4-4 u^3+2 u^2 y^3+\frac{2 y^3}{t^2}+\frac{6 y^3}{t u}+\frac{2 y^3}{u^2}+14 y^3-32 u y^2-88 u+\frac{8 u^2 y}{t^2}+24 u^2 y+\frac{6 u y}{t^3} \nn \\
    &+\frac{32 y}{t^2}+\frac{2 y}{t^4}+\frac{52 y}{t u}+\frac{6 y}{t^3 u}+\frac{24 y}{u^2}+\frac{8 y}{t^2 u^2}+\frac{2 y}{t u^3}+94 y-\frac{2 u^3}{t^2}-\frac{8 u y^2}{t^2}-\frac{32 u}{t^2}-\frac{2 u^2}{t^3}-\frac{4 y^2}{t^3}-\frac{12}{t^3} \nn \\
    &-\frac{2 u}{t^4}-\frac{2 y^4}{u}-\frac{32 y^2}{u}-\frac{88}{u}-\frac{8 y^2}{t^2 u}-\frac{32}{t^2 u}-\frac{2}{t^4 u}-\frac{8 y^2}{t u^2}-\frac{24}{t u^2}-\frac{2}{t^3 u^2}-\frac{4}{u^3}-\frac{2}{t^2 u^3}+\frac{24 u^2}{y}+\frac{94}{y} \nn \\
    &+\frac{2 u^3}{t y}+\frac{52 u}{t y}+\frac{8 u^2}{t^2 y}+\frac{32}{t^2 y}+\frac{6 u}{t^3 y}+\frac{2}{t^4 y}+\frac{52}{t u y}+\frac{6}{t^3 u y}+\frac{24}{u^2 y}+\frac{8}{t^2 u^2 y}+\frac{2}{t u^3 y}-\frac{32 u}{y^2}-\frac{8 u^2}{t y^2} \nn \\
    &-\frac{36}{t y^2}-\frac{8 u}{t^2 y^2} -\frac{4}{t^3 y^2}-\frac{32}{u y^2}-\frac{8}{t^2 u y^2}-\frac{8}{t u^2 y^2}+\frac{2 u^2}{y^3}+\frac{14}{y^3}+\frac{6 u}{t y^3}+\frac{2}{t^2 y^3}+\frac{6}{t u y^3}+\frac{2}{u^2 y^3}-\frac{2 u}{y^4} \nn \\
    &-\frac{2}{u y^4}+\frac{6 u y^3}{t}-\frac{24 u^2}{t}-\frac{8 u^2 y^2}{t}-\frac{36 y^2}{t}+\frac{2 u^3 y}{t}+\frac{52 u y}{t}-\frac{88}{t}
\end{align}
\begin{align}
    F_{200}=-2 u-\frac{2}{u}+2 y+\frac{2}{y}\;,\;\;
\end{align}
\begin{align}
    &F_{210}=-t^3 u^2-\frac{t^3}{u^2}-\frac{u^2}{t^3}-\frac{1}{t^3 u^2}+t^3 u y+\frac{t^3 y}{u}+\frac{t^3 u}{y}+\frac{t^3}{u y}+\frac{u y}{t^3}+\frac{y}{t^3 u}+\frac{u}{t^3 y}+\frac{1}{t^3 u y}-2 t^3 \nn \\
    &-\frac{2}{t^3}+t^2 u^2 y+\frac{t^2 y}{u^2}+\frac{t^2 u^2}{y}+\frac{t^2}{u^2 y}+\frac{u^2 y}{t^2}+\frac{y}{t^2 u^2}+\frac{u^2}{t^2 y}+\frac{1}{t^2 u^2 y}-t^2 u y^2-\frac{t^2 y^2}{u}-\frac{t^2 u}{y^2}-\frac{t^2}{u y^2} \nn \\
    &-\frac{u y^2}{t^2}-\frac{y^2}{t^2 u}-\frac{u}{t^2 y^2}-\frac{1}{t^2 u y^2}-3 t^2 u-\frac{3 t^2}{u}-\frac{3 u}{t^2}-\frac{3}{t^2 u}+3 t^2 y+\frac{3 t^2}{y}+\frac{3 y}{t^2}+\frac{3}{t^2 y}-2 t u^2-\frac{2 t}{u^2} \nn \\
    &-\frac{2}{t u^2}-\frac{2 u^2}{t}+4 t u y+\frac{4 t y}{u}+\frac{4 t u}{y}+\frac{4 t}{u y}+\frac{4 y}{t u}+\frac{4 u}{t y}+\frac{4}{t u y}+\frac{4 u y}{t}-2 t y^2-\frac{2 t}{y^2}-\frac{2}{t y^2}-\frac{2 y^2}{t}-8 t \nn \\
    &-\frac{8}{t}+u^2 y+\frac{y}{u^2}+\frac{u^2}{y}+\frac{1}{u^2 y}-2 u y^2-\frac{2 y^2}{u}-\frac{2 u}{y^2}-\frac{2}{u y^2}-6 u-\frac{6}{u}+y^3+\frac{1}{y^3}+7 y+\frac{7}{y}
\end{align}
\begin{align}
    &F_{211}=-t y^4-u y^4-\frac{y^4}{t}-\frac{y^4}{u}+2 t^2 y^3+2 u^2 y^3+6 t u y^3+\frac{6 u y^3}{t}+\frac{2 y^3}{t^2}+\frac{6 t y^3}{u}+\frac{6 y^3}{t u}+\frac{2 y^3}{u^2} \nn \\
    &+14 y^3-t^3 y^2-u^3 y^2-9 t u^2 y^2-33 t y^2-9 t^2 u y^2-33 u y^2-\frac{9 u^2 y^2}{t}-\frac{33 y^2}{t}-\frac{9 u y^2}{t^2}-\frac{y^2}{t^3}-\frac{9 t^2 y^2}{u} \nn \\
    &-\frac{33 y^2}{u}-\frac{9 y^2}{t^2 u}-\frac{9 t y^2}{u^2}-\frac{9 y^2}{t u^2}-\frac{y^2}{u^3}+4 t u^3 y+\frac{4 u^3 y}{t}+28 t^2 y+10 t^2 u^2 y+\frac{10 u^2 y}{t^2}+28 u^2 y+4 t^3 u y \nn \\
    &+52 t u y+\frac{52 u y}{t}+\frac{4 u y}{t^3}+\frac{28 y}{t^2}+\frac{4 t^3 y}{u}+\frac{52 t y}{u}+\frac{52 y}{t u}+\frac{4 y}{t^3 u}+\frac{10 t^2 y}{u^2}+\frac{28 y}{u^2}+\frac{10 y}{t^2 u^2}+\frac{4 t y}{u^3}+\frac{4 y}{t u^3} \nn \\
	&+90 y-8 t^3-3 t^2 u^3-8 u^3-3 t^3 u^2-29 t u^2-86 t-29 t^2 u-86 u-\frac{29 u^2}{t}-\frac{86}{t}-\frac{3 u^3}{t^2}-\frac{29 u}{t^2}\nn \\
    &-\frac{3 u^2}{t^3}-\frac{8}{t^3}-\frac{29 t^2}{u} -\frac{86}{u}-\frac{29}{t^2 u}-\frac{3 t^3}{u^2}-\frac{29 t}{u^2}-\frac{29}{t u^2}-\frac{3}{t^3 u^2}-\frac{3 t^2}{u^3}-\frac{8}{u^3}-\frac{3}{t^2 u^3}+\frac{4 u^3}{t y}+\frac{52 u}{t y} \nn \\
    &+\frac{10 u^2}{t^2 y}+\frac{28}{t^2 y}+\frac{4 u}{t^3 y}+\frac{4 t^3}{u y}+\frac{52 t}{u y}+\frac{52}{t u y}+\frac{4}{t^3 u y}+\frac{10 t^2}{u^2 y}+\frac{28}{u^2 y}+\frac{10}{t^2 u^2 y}+\frac{4 t}{u^3 y}+\frac{4}{t u^3 y}-\frac{t^3}{y^2} \nn \\
    &-\frac{u^3}{y^2}-\frac{9 t u^2}{y^2}-\frac{33 t}{y^2}-\frac{9 t^2 u}{y^2}-\frac{33 u}{y^2}-\frac{9 u^2}{t y^2}-\frac{33}{t y^2}-\frac{9 u}{t^2 y^2}-\frac{1}{t^3 y^2}-\frac{9 t^2}{u y^2}-\frac{33}{u y^2}-\frac{9}{t^2 u y^2}-\frac{9 t}{u^2 y^2} \nn \\
    &-\frac{9}{t u^2 y^2}-\frac{1}{u^3 y^2}+\frac{2 t^2}{y^3}+\frac{2 u^2}{y^3}+\frac{6 t u}{y^3}+\frac{14}{y^3}+\frac{6 u}{t y^3}+\frac{2}{t^2 y^3}+\frac{6 t}{u y^3}+\frac{6}{t u y^3}+\frac{2}{u^2 y^3}-\frac{t}{y^4}-\frac{u}{y^4}-\frac{1}{t y^4} \nn \\
    &-\frac{1}{u y^4}+\frac{4 t u^3}{y}+\frac{28 t^2}{y}+\frac{10 t^2 u^2}{y}+\frac{28 u^2}{y}+\frac{4 t^3 u}{y}+\frac{52 t u}{y}+\frac{90}{y}\;,
\end{align}
and further up to $F_{444}(t,u,y)$.

\begin{figure}[t!]
	\begin{center}
    \includegraphics[width=5cm]{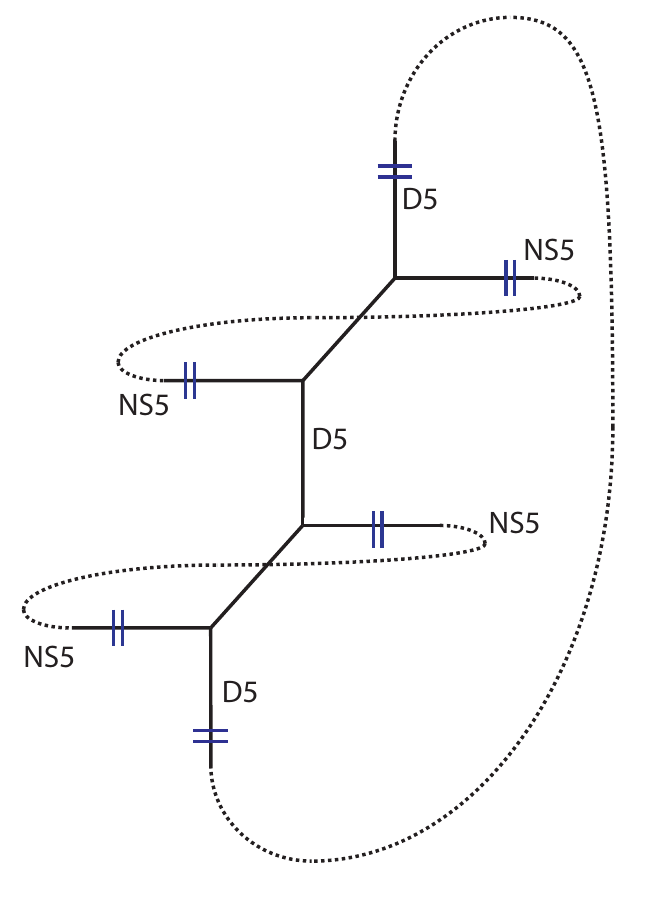}
	\caption{$(p,q)$ fivebranes web dual to rank 2 little string theory}\label{Fig:rRank2-5branes}
	\end{center}
\end{figure}

Let us define the following variables,
\begin{equation}
	\hat{w} = w y^{-1}\;,\;\;\hat{v}_{1} =v_{1}y^{-1}\;,\;\;\hat{v}_{2} =v_{2}y^{-1}
\end{equation}
We can check $\hat{w} \leftrightarrow y$ exchange symmetry of the index. 
This is an analog of the triality exchanging $(\hat{q},\hat{w},y)$ at $N=1$. 
It can be understood as a geometric duality of the dual $(p,q)$-fivebrane
web diagram Fig.\ref{Fig:rRank2-5branes} of the rank 2 little string theory.
Namely, let us define the index
\begin{equation}
  \tilde{Z}=PE[2I_{\rm com}y]Z_{\rm IIA}\ .
\end{equation}
We find that $\tilde{Z}$ is invariant under the $\hat{w} \leftrightarrow y$ exchange, 
to some high orders in fugacities.

\subsection{Three NS5-branes}

The index of $U(3)$ IIB little string theory is given by
\begin{align}
	Z_{\rm IIB}(\alpha_{i},\epsilon_{\pm},m;q,w) &=Z^{\rm IIB}_{\rm pert}(\alpha_{i},\epsilon_{\pm},m;q) Z^{\rm IIB}_{\rm inst}(\alpha_{i},\epsilon_{\pm},m;q,w)
\end{align}
with
\begin{align}
	Z^{\rm IIB}_{\rm pert}&= PE\left[
	I_{+}(v_{1}+v_{2}+v_{1}v_{2})+ 3I_{+}\frac{q}{1-q}\right] \nn \\
	&\times PE\left[I_{+}\left(v_{1}+v_{1}^{-1}+v_{2}+v_{2}^{-1}+v_{1}v_{2}+v_{1}^{-1}v_{2}^{-1}\right)\frac{q}{1-q}\right] \nn \\	 &=PE\left[I_{+}\frac{v_{1}+v_{2}+v_{3}+v_{1}v_{2}+v_{1}v_{2}+v_{2}v_{3}+3v_{1}v_{2}v_{3}}{1-v_{1}v_{2}v_{3}}\right].
\end{align}
where $v_{1}=e^{2 \pi i \alpha_{12}}$, $v_{2}=e^{2 \pi i \alpha_{23}}$, and $v_{3} = q v_{1}^{-1}v_{2}^{-1}$. 
$Z_{k}$'s appearing in $Z^{\rm IIB}_{\rm inst}$ are obtained from eq.(\ref{Instanton}).

The index of the rank 3 IIA little string theory is given by
\begin{align}
	\hat{Z}_{\rm IIA} &= Z_{\rm extra}(q)^{-1}Z^{\rm IIA}_{\rm mom}(\epsilon_{\pm},m;w)Z^{\rm IIA}_{\rm string}(\alpha_{i},\epsilon_{\pm},m;w,q) \nn \\
	&=Z_{\rm extra}(q)^{-1}Z^{\rm IIA}_{\rm mom}(\epsilon_{\pm},m;w)\sum_{n_{i}=0}^{\infty}v_{1}^{n_{1}}v_{2}^{n_{2}}v_{3}^{n_{3}}Z^{(n_{1},n_{2},n_{3})}_{\rm string}(\epsilon_{\pm},m;w)
\end{align}
$Z_{\rm extra}(q)$ is given by
\begin{equation}
	Z_{\rm extra} = PE\left[\frac{q}{1-q}\right] = PE\left[\frac{v_{1}v_{2}v_{3}}{1-v_{1}v_{2}v_{3}}\right]
\end{equation}
$Z^{\rm IIA}_{\rm mom}(\epsilon_{\pm},m;w)$ is given by
\begin{equation}
	Z^{\rm IIA}_{\rm mom}(\epsilon_{\pm},m;w)=PE\left[3I_{-}(\epsilon_{\pm},m)\frac{w}{1-w} \right]\;.
\end{equation}
Note that $Z^{(n_{1},n_{2},n_{3})}_{\rm string}$ is invariant under the permutation of
$n_{1}$, $n_{2}$ and $n_{3}$, from the symmetry of the quiver. The elliptic genera of the IIA fractional little strings, $Z^{(n_{1},n_{2},n_{3})}_{\rm string}$, are obtained from eq.(\ref{IIAstring}). For example,
\begin{equation}
	Z^{(1,0,0)}_{\rm string}(\epsilon_{\pm},m;w)=\frac{\theta_{1}(w,m \pm\epsilon_{+})}{\theta_{1}(w,\epsilon_{1})\theta_{1}(w,\epsilon_{2})}\;,\;\;
	Z^{(1,1,0)}_{\rm string}(\epsilon_{\pm},m;w)=\frac{\theta _1\left(w;m \pm\epsilon _-\right)\theta _1\left(w;m\pm\epsilon _+\right)}{\theta _1\left(w;\epsilon _1\right){}^2 \theta _1\left(w;\epsilon _2\right){}^2}
\end{equation}
\begin{align}
	&Z^{(2,0,0)}_{\rm string}(\epsilon_{\pm},m;w)=
	Z^{(1,0,0)}_{\rm string} \cdot \left(\frac{ \theta _1\left(w;\epsilon _++\epsilon _1\pm m\right)}{\theta _1\left(w;2 \epsilon _1\right) \theta _1\left(w;\epsilon _1-\epsilon _2\right)}-(\epsilon_{1}\leftrightarrow \epsilon_{2})\right)
\end{align}
\begin{equation}
	Z^{(1,1,1)}_{\rm string}(\epsilon_{\pm},m;w)=\frac{\theta_{1}(w,m\pm \epsilon_{-})^{3}}{\theta_{1}(w,\epsilon_{1})^{3}\theta_{1}(w,\epsilon_{2})^{3}}
\end{equation}

We write the IIB/IIA indices as
\begin{equation}
	Z_{\rm IIB}(\alpha_{i},\epsilon_{\pm},m;w,v_{i}) = PE\left[I_{\rm com}(t,u)\sum_{i,j,k,l=0}^{\infty}F^{\rm IIB}_{ijkl}(t,u,y)w^{i}v_{1}^{j}v_{2}^{k}v_{3}^{l}\right]\;,
\end{equation}
\begin{equation}
	\hat{Z}_{\rm IIA}(\alpha_{i},\epsilon_{\pm},m;w,v_{i})=PE\left[I_{\rm com}(t,u)\sum_{i,j,k,l=0}^{\infty}F^{\rm IIA}_{ijkl}(t,u,y)w^{i}v_{1}^{j}v_{2}^{k}v_{3}^{l}\right]\;,
\end{equation}
where $I_{\rm com}$ is given by eq.(\ref{Icom}). The coefficients $F_{ijkl}(t,u,y)$ are 
polynomials of $t$, $u$, and $y$, satisfying $F_{ijkl}=F_{i(jkl)}$.

T-duality implies that $F^{\rm IIA}_{ijkl}=F^{\rm IIB}_{ijkl}\equiv F_{ijkl}$. 
We checked this for higher orders of the fugacity variables. 
$F_{ijkl}$ are given by
\begin{equation}
	F_{0100}=-t-\frac{1}{t}+y+\frac{1}{y}\;,\;\;F_{0200}=0\;,\;\;
F_{0110}=-t-\frac{1}{t}+y+\frac{1}{y}\;,\;\;F_{0210}=0
\end{equation}
\begin{equation}
	F_{0111}=-3 t-\frac{3}{t}+3 y+\frac{3}{y}\;,\;\;F_{0220}=0\;,\;\;
F_{0211}=-t-\frac{1}{t}+y+\frac{1}{y}
\end{equation}
\begin{equation}
	F_{0221}=-t-\frac{1}{t}+y+\frac{1}{y}\;,\;\;F_{0222}=-3 t-\frac{3}{t}+3 y+\frac{3}{y}\;,\;\;F_{1000}=-3 u-\frac{3}{u}+3 y+\frac{3}{y}
\end{equation}
\begin{align}
	F_{1100}=&-t^2u-\frac{t^2}{u}-\frac{u}{t^2}-\frac{1}{t^2 u}+t^2 y+\frac{t^2}{y}+\frac{y}{t^2}+\frac{1}{t^2 y}+t u y+\frac{t y}{u}+\frac{t u}{y}+\frac{t}{u y}+\frac{y}{t u}+\frac{u}{t y}+\frac{1}{t u y}\nn \\
	&+\frac{u y}{t}-t y^2 -\frac{t}{y^2}-\frac{1}{t y^2}-\frac{y^2}{t}-2 t-\frac{2}{t}-2 u-\frac{2}{u}+2 y+\frac{2}{y}
\end{align}
\begin{align}
	F_{1200}=&-t^4u-\frac{t^4}{u}-\frac{u}{t^4}-\frac{1}{t^4 u}+t^4 y+\frac{t^4}{y}+\frac{y}{t^4}+\frac{1}{t^4 y}+t^3 u y+\frac{t^3 y}{u}+\frac{t^3 u}{y}+\frac{t^3}{u y}+\frac{u y}{t^3}+\frac{y}{t^3 u}\nn \\
	&+\frac{u}{t^3 y}+\frac{1}{t^3 u y}-t^3 y^2-\frac{t^3}{y^2} -\frac{y^2}{t^3}-\frac{1}{t^3 y^2}-2 t^3-\frac{2}{t^3}-2 t^2 u-\frac{2 t^2}{u}-\frac{2 u}{t^2}-\frac{2}{t^2 u}+2 t^2 y\nn \\
	&+\frac{2 t^2}{y} +\frac{2 y}{t^2}+\frac{2}{t^2 y}+t u y+\frac{t y}{u}+\frac{t u}{y}+\frac{t}{u y}+\frac{y}{t u}+\frac{u}{t y}+\frac{1}{t u y}+\frac{u y}{t}-t y^2-\frac{t}{y^2}-\frac{1}{t y^2} \nn \\
	&-\frac{y^2}{t}-2 t-\frac{2}{t}-2 u-\frac{2}{u}+2 y+\frac{2}{y}
\end{align}
\begin{align}
    F_{1110}=&-2 t^2 u-\frac{2 u}{t^2}-\frac{2 t^2}{u}-\frac{2}{t^2 u}+2 t^2 y+\frac{2 y}{t^2}+\frac{2}{t^2 y}+\frac{2 t^2}{y}+3 t u y+\frac{3 u y}{t}+\frac{3 t y}{u}+\frac{3 y}{t u}+\frac{3 u}{t y} \nn\\
    &+\frac{3 t}{u y}+\frac{3}{t u y}+\frac{3 t u}{y}-3 t y^2-\frac{3 y^2}{t}-\frac{3 t}{y^2}-\frac{3}{t y^2}-6 t-\frac{6}{t}-u y^2-\frac{y^2}{u}-\frac{u}{y^2}-\frac{1}{u y^2}-6 u \nn \\
    &-\frac{6}{u}+y^3+\frac{1}{y^3}+7 y+\frac{7}{y}
\end{align}
\begin{align}
	F_{1210}=& -t^4u-\frac{t^4}{u}-\frac{u}{t^4}-\frac{1}{t^4 u}+t^4 y+\frac{t^4}{y}+\frac{y}{t^4}+\frac{1}{t^4 y}+2 t^3 u y+\frac{2 t^3 y}{u}+\frac{2 t^3 u}{y}+\frac{2 t^3}{u y}+\frac{2 u y}{t^3}+\frac{2 y}{t^3 u} \nn \\
	&+\frac{2 u}{t^3 y}+\frac{2}{t^3 u y}-2 t^3 y^2-\frac{2 t^3}{y^2}-\frac{2 y^2}{t^3}-\frac{2}{t^3 y^2}-4 t^3-\frac{4}{t^3}-t^2 u y^2-\frac{t^2 y^2}{u}-\frac{t^2 u}{y^2}-\frac{t^2}{u y^2}-\frac{u y^2}{t^2} \nn \\
	&-\frac{y^2}{t^2 u}-\frac{u}{t^2 y^2}-\frac{1}{t^2 u y^2}-5 t^2 u-\frac{5 t^2}{u}-\frac{5 u}{t^2}-\frac{5}{t^2 u}+t^2 y^3+\frac{t^2}{y^3}+\frac{y^3}{t^2}+\frac{1}{t^2 y^3}+6 t^2 y+\frac{6 t^2}{y} \nn \\
	&+\frac{6 y}{t^2}+\frac{6}{t^2 y}+4 t u y+\frac{4 t y}{u}+\frac{4 t u}{y}+\frac{4 t}{u y}+\frac{4 y}{t u}+\frac{4 u}{t y}+\frac{4}{t u y}+\frac{4 u y}{t}-4 t y^2-\frac{4 t}{y^2}-\frac{4}{t y^2}-\frac{4 y^2}{t} \nn \\
	&-8 t-\frac{8}{t}-u y^2-\frac{y^2}{u}-\frac{u}{y^2}-\frac{1}{u y^2}-6 u-\frac{6}{u}+y^3+\frac{1}{y^3}+7 y+\frac{7}{y}
\end{align}
\begin{align}
	F_{1111} =& -6 t^2 u-\frac{6 u}{t^2}-\frac{6 t^2}{u}-\frac{6}{t^2 u}+6 t^2 y+\frac{6 y}{t^2}+\frac{6}{t^2 y}+\frac{6 t^2}{y}-3 t u^2-\frac{3 u^2}{t}-\frac{3 t}{u^2}-\frac{3}{t u^2}+15 t u y\nn \\
	&+\frac{15 u y}{t} +\frac{15 t y}{u}+\frac{15 y}{t u}+\frac{15 u}{t y}+\frac{15 t}{u y}+\frac{15}{t u y}+\frac{15 t u}{y}-12 t y^2-\frac{12 y^2}{t}-\frac{12 t}{y^2}-\frac{12}{t y^2}-30 t\nn \\
	&-\frac{30}{t}+3 u^2 y+\frac{3 y}{u^2}+\frac{3}{u^2 y}+\frac{3 u^2}{y}-9 u y^2-\frac{9 y^2}{u}-\frac{9 u}{y^2}-\frac{9}{u y^2}-30 u-\frac{30}{u}+6 y^3+\frac{6}{y^3} \nn \\
	&+36 y +\frac{36}{y}
\end{align}
\begin{align}
    F_{1220}&=-2 t^4 u-\frac{2 t^4}{u}-\frac{2 u}{t^4}-\frac{2}{t^4 u}+2 t^4 y+\frac{2 t^4}{y}+\frac{2 y}{t^4}+\frac{2}{t^4 y}+3 t^3 u y+\frac{3 t^3 y}{u}+\frac{3 t^3 u}{y}+\frac{3 t^3}{u y}+\frac{3 u y}{t^3}\nn \\
    &+\frac{3 y}{t^3 u}+\frac{3 u}{t^3 y} +\frac{3}{t^3 u y}-3 t^3 y^2-\frac{3 t^3}{y^2}-\frac{3 y^2}{t^3}-\frac{3}{t^3 y^2}-6 t^3-\frac{6}{t^3}-t^2 u y^2-\frac{t^2 y^2}{u}-\frac{t^2 u}{y^2}-\frac{t^2}{u y^2}\nn \\
    &-\frac{u y^2}{t^2}-\frac{y^2}{t^2 u} -\frac{u}{t^2 y^2}-\frac{1}{t^2 u y^2}-8 t^2 u-\frac{8 t^2}{u}-\frac{8 u}{t^2}-\frac{8}{t^2 u}+t^2 y^3+\frac{t^2}{y^3}+\frac{y^3}{t^2}+\frac{1}{t^2 y^3} +9 t^2 y\nn \\
    &+\frac{9 t^2}{y}+\frac{9 y}{t^2}+\frac{9}{t^2 y}+7 t u y+\frac{7 t y}{u}+\frac{7 t u}{y}+\frac{7 t}{u y}+\frac{7 y}{t u}+\frac{7 u}{t y}+\frac{7}{t u y}+\frac{7 u y}{t}-7 t y^2-\frac{7 t}{y^2}-\frac{7}{t y^2} \nn \\
    &-\frac{7 y^2}{t}-14 t-\frac{14}{t}-2 u y^2-\frac{2 y^2}{u}-\frac{2 u}{y^2}-\frac{2}{u y^2}-12 u-\frac{12}{u}+2 y^3+\frac{2}{y^3}+14 y+\frac{14}{y}
\end{align}
\begin{align}
    F_{1211}&=-2 u t^4+2 y t^4-\frac{2 t^4}{u}+\frac{2 t^4}{y}-u^2 t^3-5 y^2 t^3+6 u y t^3+\frac{6 y t^3}{u}-\frac{t^3}{u^2}+\frac{6 u t^3}{y}+\frac{6 t^3}{u y}-\frac{5 t^3}{y^2} \nn \\
    &-12 t^3+4 y^3 t^2-6 u y^2 t^2-22 u t^2+2 u^2 y t^2+\frac{2 y t^2}{u^2}+26 y t^2-\frac{6 y^2 t^2}{u}-\frac{22 t^2}{u}+\frac{2 u^2 t^2}{y}+\frac{26 t^2}{y} \nn \\
    &+\frac{2 t^2}{u^2 y}-\frac{6 u t^2}{y^2}-\frac{6 t^2}{u y^2}+\frac{4 t^2}{y^3}-y^4 t+2 u y^3 t+\frac{2 y^3 t}{u}-5 u^2 t-u^2 y^2 t-24 y^2 t+27 u y t+\frac{27 y t}{u}\nn \\
    &-\frac{y^2 t}{u^2}-\frac{5 t}{u^2} +\frac{27 u t}{y}+\frac{27 t}{u y}-\frac{u^2 t}{y^2}-\frac{24 t}{y^2}-\frac{t}{u^2 y^2}+\frac{2 u t}{y^3}+\frac{2 t}{u y^3}-\frac{t}{y^4}-52 t+\frac{4 y^3}{t^2}+\frac{2 y^3}{t u} \nn \\
    &+9 y^3-13 u y^2-42 u+\frac{2 u^2 y}{t^2}+4 u^2 y+\frac{6 u y}{t^3}+\frac{26 y}{t^2}+\frac{2 y}{t^4}+\frac{27 y}{t u}+\frac{6 y}{t^3 u}+\frac{4 y}{u^2}+\frac{2 y}{t^2 u^2}+51 y\nn \\
    &-\frac{6 u y^2}{t^2}-\frac{22 u}{t^2}-\frac{u^2}{t^3}-\frac{5 y^2}{t^3}-\frac{12}{t^3}-\frac{2 u}{t^4}-\frac{13 y^2}{u}-\frac{42}{u}-\frac{6 y^2}{t^2 u}-\frac{22}{t^2 u}-\frac{2}{t^4 u}-\frac{y^2}{t u^2}-\frac{5}{t u^2} \nn \\
    &-\frac{1}{t^3 u^2}+\frac{4 u^2}{y}+\frac{51}{y}+\frac{27 u}{t y}+\frac{2 u^2}{t^2 y}+\frac{26}{t^2 y}+\frac{6 u}{t^3 y}+\frac{2}{t^4 y}+\frac{27}{t u y}+\frac{6}{t^3 u y}+\frac{4}{u^2 y}+\frac{2}{t^2 u^2 y}-\frac{13 u}{y^2} \nn \\
    &-\frac{u^2}{t y^2}-\frac{24}{t y^2}-\frac{6 u}{t^2 y^2}-\frac{5}{t^3 y^2}-\frac{13}{u y^2}-\frac{6}{t^2 u y^2}-\frac{1}{t u^2 y^2}+\frac{9}{y^3}+\frac{2 u}{t y^3}+\frac{4}{t^2 y^3}+\frac{2}{t u y^3}-\frac{1}{t y^4} \nn \\
    &-\frac{y^4}{t}+\frac{2 u y^3}{t}-\frac{5 u^2}{t}-\frac{u^2 y^2}{t}-\frac{24 y^2}{t}+\frac{27 u y}{t}-\frac{52}{t}
\end{align}
\begin{align}
    F_{2000}=-3 u-\frac{3}{u}+3 y+\frac{3}{y}
\end{align}
\begin{align}
    F_{2100}&=-t^3 u^2-\frac{t^3}{u^2}-\frac{u^2}{t^3}-\frac{1}{t^3 u^2}+t^3 u y+\frac{t^3 y}{u}+\frac{t^3 u}{y}+\frac{t^3}{u y}+\frac{u y}{t^3}+\frac{y}{t^3 u}+\frac{u}{t^3 y}+\frac{1}{t^3 u y}-2 t^3 \nn \\
    &-\frac{2}{t^3}+t^2 u^2 y+\frac{t^2 y}{u^2}+\frac{t^2 u^2}{y}+\frac{t^2}{u^2 y}+\frac{u^2 y}{t^2}+\frac{y}{t^2 u^2}+\frac{u^2}{t^2 y}+\frac{1}{t^2 u^2 y}-t^2 u y^2-\frac{t^2 y^2}{u} - \frac{t^2 u}{y^2}\nn \\
    &-\frac{t^2}{u y^2}-\frac{u y^2}{t^2}-\frac{y^2}{t^2 u}-\frac{u}{t^2 y^2}-\frac{1}{t^2 u y^2}-3 t^2 u-\frac{3 t^2}{u}-\frac{3 u}{t^2}-\frac{3}{t^2 u}+3 t^2 y+\frac{3 t^2}{y}+\frac{3 y}{t^2}+\frac{3}{t^2 y} \nn \\
    &-2 t u^2-\frac{2 t}{u^2}-\frac{2}{t u^2}-\frac{2 u^2}{t}+4 t u y+\frac{4 t y}{u}+\frac{4 t u}{y}+\frac{4 t}{u y}+\frac{4 y}{t u}+\frac{4 u}{t y}+\frac{4}{t u y}+\frac{4 u y}{t}-2 t y^2-\frac{2 t}{y^2} \nn \\
    &-\frac{2}{t y^2}-\frac{2 y^2}{t}-8 t-\frac{8}{t}+u^2 y+\frac{y}{u^2}+\frac{u^2}{y}+\frac{1}{u^2 y}-2 u y^2-\frac{2 y^2}{u}-\frac{2 u}{y^2}-\frac{2}{u y^2}-6 u-\frac{6}{u} \nn \\
    &+y^3+\frac{1}{y^3}+7 y+\frac{7}{y}
\end{align}
\begin{align}
    F_{2110}&=-t y^4-\frac{y^4}{t}+2 t^2 y^3+3 t u y^3+\frac{3 u y^3}{t}+\frac{2 y^3}{t^2}+\frac{3 t y^3}{u}+\frac{3 y^3}{t u}+7 y^3-t^3 y^2-2 t u^2 y^2-17 t y^2 \nn \\
    &-7 t^2 u y^2-16 u y^2-\frac{2 u^2 y^2}{t}-\frac{17 y^2}{t}-\frac{7 u y^2}{t^2}-\frac{y^2}{t^3}-\frac{7 t^2 y^2}{u}-\frac{16 y^2}{u}-\frac{7 y^2}{t^2 u}-\frac{2 t y^2}{u^2}-\frac{2 y^2}{t u^2}\nn \\
    &+19 t^2 y+5 t^2 u^2 y+\frac{5 u^2 y}{t^2}+9 u^2 y+4 t^3 u y+26 t u y +\frac{26 u y}{t}+\frac{4 u y}{t^3}+\frac{19 y}{t^2}+\frac{4 t^3 y}{u}+\frac{26 t y}{u}\nn \\
    &+\frac{26 y}{t u}+\frac{4 y}{t^3 u}+\frac{5 t^2 y}{u^2}+\frac{9 y}{u^2}+\frac{5 y}{t^2 u^2}+45 y-8 t^3-3 t^3 u^2-12 t u^2-48 t-17 t^2 u-38 u -\frac{12 u^2}{t} \nn \\
    &-\frac{48}{t}-\frac{17 u}{t^2}-\frac{3 u^2}{t^3}-\frac{8}{t^3}-\frac{17 t^2}{u}-\frac{38}{u}-\frac{17}{t^2 u}-\frac{3 t^3}{u^2}-\frac{12 t}{u^2}-\frac{12}{t u^2}-\frac{3}{t^3 u^2}+\frac{26 u}{t y}+\frac{5 u^2}{t^2 y} +\frac{19}{t^2 y}\nn \\
    &+\frac{4 u}{t^3 y}+\frac{4 t^3}{u y}+\frac{26 t}{u y}+\frac{26}{t u y}+\frac{4}{t^3 u y}+\frac{5 t^2}{u^2 y}+\frac{9}{u^2 y}+\frac{5}{t^2 u^2 y}-\frac{t^3}{y^2}-\frac{2 t u^2}{y^2}-\frac{17 t}{y^2}-\frac{7 t^2 u}{y^2} -\frac{16 u}{y^2}\nn \\
    &-\frac{2 u^2}{t y^2}-\frac{17}{t y^2}-\frac{7 u}{t^2 y^2}-\frac{1}{t^3 y^2}-\frac{7 t^2}{u y^2}-\frac{16}{u y^2}-\frac{7}{t^2 u y^2}-\frac{2 t}{u^2 y^2}-\frac{2}{t u^2 y^2}+\frac{2 t^2}{y^3}+\frac{3 t u}{y^3}+\frac{7}{y^3}+\frac{3 u}{t y^3} \nn \\
    &+\frac{2}{t^2 y^3}+\frac{3 t}{u y^3}+\frac{3}{t u y^3}-\frac{t}{y^4}-\frac{1}{t y^4}+\frac{19 t^2}{y}+\frac{5 t^2 u^2}{y}+\frac{9 u^2}{y}+\frac{4 t^3 u}{y}+\frac{26 t u}{y}+\frac{45}{y}
\end{align}
and so on.

For the rank 3 indices, we can also check the duality of
exchanging $\hat{w}=w y^{-1}$ and $y$.

\section{$SL(2,Z)$ transformations of the elliptic genus}

The index of the winding IIB little strings is given by
\begin{equation}
 	Z^{\rm IIB}_{\rm inst}(\alpha_{i},\epsilon_{\pm},m;q,w) = \sum_{k=1}^{\infty}w^{k}\sum_{Y :\sum_{i}|Y_{i}|=k}\prod_{i,j=1}^{N}\prod_{s \in Y_{i}}\frac{\theta_{1}\left(q;E_{ij}+m- \epsilon_{-}\right)\theta_{1}\left(q;E_{ij} - m - \epsilon_{-}\right)}{\theta_{1}\left(q;E_{ij}-\epsilon_{1}\right)\theta_{1}\left(q;E_{ij} + \epsilon_2\right)}\;
\end{equation}
where
\begin{equation}
	E_{ij} = \alpha_{i}-\alpha_{j}-\epsilon_{1}h_{i}(s)+\epsilon_{2}v_{j}(s)\;.
\end{equation}
$q$ and $w$ are given by
\begin{equation}
	q = e^{2 \pi i \tau}\;,\;\;w = e^{2 \pi i \rho}\ ,
\end{equation}
where $\tau =i \frac{R_{\beta}}{R_{\rm IIB}}$ is the complex structure on the torus, 
$\alpha'\rho = i R_{\beta}R_{\rm IIB}$ is the K\"ahler parameter of it, and 
$R_\beta$ is the radius of the temporal circle.

The modular transformation of the Jacobi's theta function is given by
\begin{equation}
	\theta_{1}\left(-\frac{1}{\tau}; \frac{z}{\tau}\right) = -i(-i \tau^{\frac{1}{2}})\exp\left(\frac{i \pi z^{2}}{\tau}\right)\theta_{1}(\tau;z)
\end{equation}
Using this property, the S-duality transformation of $Z^{\rm IIB}_{\rm inst}$
in $\tau$ is given by
\begin{equation}
	\sum_{k=1}^{\infty}w^{k}\exp\left(-2 \pi i\frac{m^{2}-\epsilon_{+}^{2}}{\tau}kN\right)\sum_{Y :\sum_{i}|Y_{i}|=k}\prod_{i,j=1}^{N}\prod_{s \in Y_{i}}\frac{\theta_{1}\left(-\frac{1}{\tau};\frac{E_{ij}+m- \epsilon_{-}}{ \tau}\right)\theta_{1}\left(-\frac{1}{\tau};\frac{E_{ij} - m - \epsilon_{-}}{\tau}\right)}{\theta_{1}\left(-\frac{1}{\tau};\frac{E_{ij}-\epsilon_{1}}{\tau}\right)\theta_{1}\left(-\frac{1}{\tau};\frac{E_{ij} + \epsilon_2}{\tau}\right)}\;.
\end{equation}
Transforming the fugacity variable for the winding number, $w$ by
\begin{equation}
	w \rightarrow \tilde{w}=e^{-2 \pi i\frac{m^{2}-\epsilon_{+}^{2}}{\tau}N}w\;,
\end{equation}
$Z_{\rm inst}^{\rm IIB}$ is invariant under the following transformation,
\begin{equation}
	q=e^{2 \pi i \tau} \rightarrow \tilde{q} = e^{ -\frac{2 \pi i}{\tau} }\;,\;\;w \rightarrow \tilde{w}=e^{-2 \pi i\frac{m^{2}-\epsilon_{+}^{2}}{\tau}N}w\;.
\end{equation}

The elliptic genus of the IIA strings is given by
\begin{align}
    Z^{\rm IIA}_{\rm string}(\alpha_{i},\epsilon_{\pm},m;q',w')  &=\sum_{n_{i}=0}^{\infty}e^{2 \pi i \sum_{i=1}^{N}n_{i} \alpha_{i,i+1}}Z^{(n_{1},...,n_{N})}_{\rm string}(\epsilon_{\pm},m;q') \nn \\
    &=\sum_{n_{i}=0}^{\infty}(v_{1})^{n_{1}}\cdots(v_{N})^{n_{N}}Z^{(n_{1},...,n_{N})}_{\rm string}(\epsilon_{\pm},m;q') \nn \\
    &=\sum_{n_{i}=0}^{\infty}(v_{1})^{n_{1}-n_{N}}\cdots(v_{N-1})^{n_{N-1}-n_{N}}(w')^{n_{N}}Z^{(n_{1},...,n_{N})}_{\rm string}(\epsilon_{\pm},m;q')
\end{align}
where $v_{i} = e^{ 2 \pi i (\alpha_{i,i+1})}$ and $e^{-2 \pi i(\alpha_{N+1})} = e^{-2 \pi i \alpha_{1}}w'$. $Z^{(n_{1},...,n_{N})}_{\rm string}$ is given by
\begin{align}
    Z^{(n_{1},...,n_{N})}_{\rm string}(\epsilon_{\pm},m;q')   &=\sum_{\{Y_{1},\cdots,Y_{N}\};|Y_{i}|=n_{i}}\prod_{i=1}^{N}\prod_{(a,b)\in Y_{i}}\frac{\theta_{1}(q';E_{i,i+1}^{(a,b)}-m+\epsilon_{-})\theta_{1}(q';E_{i,i-1}^{(a,b)}+m+\epsilon_{-})}{\theta_{1}(q';E_{i,i}^{(a,b)}+ \epsilon_{1})\theta_{1}(q';E_{i,i}^{(a,b)}-\epsilon_{2})}\;,
\end{align}
where
\begin{equation}
    E_{ij}^{(a,b)}=(Y_{i,a}-b)\epsilon_{1}-(Y^{T}_{j,b}-a)\epsilon_{2}\;.
\end{equation}
Upon T-duality transformation, the complex structure and the K\"ahler 
parameter are exchanged. The modular transformation of 
$Z^{(n_{1},...,n_{N})}_{\rm string}$ in $\rho\equiv\tau^\prime$ is given by
\begin{align}
    &Z^{(n_{1},\cdots,n_{N})}_{\rm string}(\epsilon_{\pm},m;\rho) \nn \\
    &\;\;\;\;=\exp\left[-\frac{\pi i }{\rho}\left(\epsilon_{1}\epsilon_{2}\sum_{a=1}^{N}(n_{a}-n_{a+1})^{2}+2(m^{2}-\epsilon_{+}^{2})\sum_{a=1}^{N}n_{a} \right)\right] \cdot Z^{(n_{1},\cdots,n_{N})}_{\mathrm{string}}\left(\frac{\epsilon_{\pm}}{\rho},\frac{m}{\rho};-\frac{1}{\rho}\right) \;.
\end{align}
where $n_{N+1} = n_{1}$. Via T-duality relation, this would imply a definite S-duality
transformation of $Z^{\rm IIB}$ in $\rho$, which would have been difficult to obtain
directly without knowing the T-dual expression.
Note that the above S-duality transformation becomes paraticularly simpler
when $n_1=n_2=\cdots=n_N$:
\begin{equation}
    Z^{(n,\cdots,n)}_{\rm string}(\epsilon_{\pm},m;\rho)
    =\exp\left[-\frac{2\pi i }{\rho}(m^{2}-\epsilon_{+}^{2})Nn\right] \cdot Z^{(n_{1},\cdots,n_{N})}_{\mathrm{string}}\left(\frac{\epsilon_{\pm}}{\rho},\frac{m}{\rho};-\frac{1}{\rho}\right) \;.
\end{equation}
The prefactor can be absorbed into a scaling of $w^\prime=q$ fugacity,
conjugate to $n$. This expression might be useful to understand the DLCQ of
type IIB little strings, in which $U(n)^N$ gauge theory description was
used \cite{Aharony:1999dw}.

\section{Concluding remarks}

In this paper, we explored the 2 dimensional $\mathcal{N}=(4,4)$ and
$\mathcal{N}=(0,4)$ gauge theory descriptions of macroscopic IIA/IIB
little strings. In particular, we proposed a new $(0,4)$ gauge theory
which enables the computation of the IIA strings' elliptic genera. We
used these elliptic genera to study the little string T-duality.

The elliptic genus is enjoying $SL(2,\mathbb{Z})\times SL(2,\mathbb{Z})$ symmetry on
the complex structure $\tau$ and Kahler parameter $\rho$ of the torus.
Interesting extended dualities were studied in \cite{Hollowood:2003cv}
for 6d maximal SYM theory compactified on $T^2$, from its
Seiberg-Witten curve. It will be interesting to see whether a larger
duality than what we explored here is realized in the elliptic genera.

It will also be interesting to see if one can study the T-duality of 
elliptic genera for the heterotic little string theories, living
on the heterotic 5-branes in the $SO(32)$ and $E_8\times E_8$ theories.
Just like our IIA strings are closely related to the `M-strings'
of 6d $(2,0)$ CFT, the $E_8\times E_8$ little strings would be
closely related to the so-called E-strings of the 6d $(1,0)$ CFT,
with $E_8$ global symmetry \cite{Witten:1996qb,Klemm:1996hh}. 
The E-string elliptic genera have been
recently studied in \cite{Kim:2014dza}, from 2d $(0,4)$ gauge theories.

Finaly, the self-dual string elliptic genera in 6d SCFTs turn out to be 
related to other interesting observables,
such as the superconformal indices \cite{Kim:2012ava}. It will be interesting to
see if the elliptic genera for little strings also find similar 
interesting applications.

\vskip 0.5cm

\hspace*{-0.8cm} {\bf\large Acknowledgements}
\vskip 0.2cm

\hspace*{-0.75cm}
This work is supported in part by the National Research Foundation of Korea (NRF)
Grants No. 2012R1A1A2042474 (JK,SK), 2012R1A2A2A02046739 (SK),
2006-0093850 (KL), 2009-0084601 (KL).

\appendix

\section{$\mathcal{N}=(4,4)$ gauge theory of IIB strings}
\label{Appendix:2d-Action}

The Lagrangian of the 2d $N=(4,4)$ gauge theory for IIB strings is given by
\begin{equation}
	\mathcal{L} = \mathcal{L}_{1}+ \mathcal{L}_{2}\;.
\end{equation}
$\mathcal{L}_{1}$ is given by
\begin{align}
    \mathcal{L}_{1} =& \frac{1}{g_{QM}^{2}}\Tr\left[-\frac{1}{4}(F_{\mu \nu})^{2}-\frac{1}{2}(D_{\mu}\varphi_{aA})(D^{\mu}\varphi^{Aa})-\frac{1}{2}(D_{\mu}a_{\alpha \dot\beta})(D^{\mu}a^{\dot{\beta}\alpha}) + \frac{1}{2}[a_{\alpha \dot{\beta}},\varphi_{a A}]^{2} \right. \nn \\
    & + \frac{i}{2}(\bar{\lambda}_{a}^{\dot{\alpha}})^{\dag}(D_{t}+D_{s})\bar{\lambda}^{\dot\alpha}_{a} + \frac{i}{2}(\bar{\lambda}^{A \dot{\alpha}})^{\dag}(D_{t}-D_{s})\bar{\lambda}^{A \dot{\alpha}}  + \frac{i}{2}(\lambda^{A}_{\alpha})^{\dag}(D_{t}+D_{s})\lambda^{A}_{\alpha}  + \frac{i}{2}(\lambda_{a \alpha})^{\dag}(D_{t}-D_{s})\lambda_{a \alpha} \nn \\
    &+ \frac{1}{2}D^{I}D^{I} - D^{I}\Big(\bar{q}^{\dot{\alpha}}q_{\dot{\beta}}(\tau^{I})^{\dot\alpha}{}_{\dot\beta}+\frac{1}{2}(\tau^{I})^{\dot{\alpha}}{}_{\dot{\beta}}[a^{\dot{\beta}\alpha},a_{\alpha \dot{\alpha}}]-\zeta^{I}\Big) + \frac{1}{2}D^{I'}D^{I'} - D^{I'}\Big(%\bar{q}^{A}q_{B}(\tau^{I'})^{B}{}_{A} + %
    \frac{1}{2}(\tau^{I'})^{A}{}_{B}[\varphi^{Ba},\varphi_{a A}]%-\zeta^{I'}%
    \Big) \nn \\
    & -\frac{i}{\sqrt{2}}(\lambda_{a \alpha})^{\dag}[a_{\alpha \dot{\beta}},\bar{\lambda}^{\dot{\beta}}_{a}]-\frac{i}{\sqrt{2}}(\lambda^{A}_{\alpha})^{\dag}[a_{\alpha \dot{\beta}},\bar{\lambda}^{A \dot{\beta}}] + \frac{i}{\sqrt{2}}(\bar\lambda^{A \dot{\alpha}})^{\dag}[a^{\dot\alpha \beta},\lambda^{A}_{\beta}] + \frac{i}{\sqrt{2}}(\bar\lambda_{a}^{\dot{\alpha}})^{\dag}[a^{\dot\alpha \beta},\lambda_{a \beta}] \nn \\
    & \left.+ \frac{i}{\sqrt{2}}(\bar{\lambda}^{\dot\alpha}_{a})^{\dag}[\varphi_{a A}, \bar{\lambda}^{A \dot{\alpha}}] + \frac{i}{\sqrt{2}}(\bar{\lambda}^{A \dot\alpha})^{\dag}[\varphi^{Aa}, \bar{\lambda}^{\dot{\alpha}}_{a}] - \frac{i}{\sqrt{2}}(\lambda_{a \alpha})^{\dag}[\varphi_{aA},\lambda^{A}_{\alpha}]-\frac{i}{\sqrt{2}}(\lambda^{A}_{\alpha})^{\dag}[\varphi^{Aa},\lambda_{a \alpha}] \right] .\nn \\
\end{align}
$\mathcal{L}_{2}$ is given by,
\begin{align}
	\mathcal{L}_{2} =& \Tr\left[-D_{\mu}\bar{q}^{\dot\alpha}D_{\mu}q_{\dot{\alpha}} - \varphi_{aA} \bar{q}^{\dot{\alpha}}q_{\dot{\alpha}}\varphi^{Aa}+ i (\psi_{a})^{\dag}(D_{t}-D_{s})\psi_{a}+ i (\psi^{A})^{\dag}(D_{t}+D_{s})\psi^{A}\right. \nn \\
	&+\sqrt{2}(\psi_{a})^{\dag}(\psi^{A}\varphi_{aA}) +\sqrt{2}(\psi^{A})^{\dag}(\psi_{a}\varphi^{Aa}) + i\sqrt{2}(\lambda^{\dot{\alpha}}_{a})^{\dag}\bar{q}^{\dot{\alpha}}\psi_{a} +i\sqrt{2} (\bar{\lambda}^{A\dot{\alpha}})^{\dag}\bar{q}^{\dot{\alpha}}\psi^{A} \nn \\
	&\left. -i\sqrt{2}(\psi_{a})^{\dag}q_{\dot{\alpha}}\bar{\lambda}^{\dot{\alpha}}_{a}-i\sqrt{2}(\psi^{A})^{\dag}q_{\dot{\alpha}}\bar{\lambda}^{A\dot{\alpha}}\Big)  \right]\;.
\end{align}
In the Higgs branch, the theory describes IIB strings bound to the NS5-branes, 
whose target space is the $k$ instanton moduli space.

The reality condition of the scalar fields is given by
\begin{align}
    a_{\alpha \dot{\alpha}}= \frac{1}{\sqrt{2}}(\sigma^{m})_{\alpha \dot{\alpha}}a_{m}\;,\;\;a^{\dot{\alpha} \alpha} = \frac{1}{\sqrt{2}}(\bar{\sigma}^{m})^{\dot{\alpha} \alpha}a_{m}\;,\;\;a^{\dot{\alpha}\alpha} = \epsilon^{\alpha \beta}\epsilon^{\dot{\alpha}\dot{\beta}}a_{\beta \dot{\beta}} = (a_{\alpha \dot{\alpha}})^{\dag}\;,
\end{align}
\begin{align}
    \varphi_{a A}= \frac{1}{\sqrt{2}}(\sigma^{I})_{a A}\varphi_{I}\;,\;\;\varphi^{Aa} = \frac{1}{\sqrt{2}}(\bar{\sigma}^{I})^{Aa}\varphi_{I}\;,\;\;\varphi^{Aa} = \epsilon^{ab}\epsilon^{AB}\varphi_{bB} = (\varphi_{bB})^{\dag}\;,
\end{align}
with $m=1,2,3,4$ and $I=1,2,3,4$. The fermions satisfy the following 
reality conditions,
\begin{equation}
	\lambda_{a \alpha} = -\epsilon_{\alpha \beta}\epsilon_{a b}(\lambda_{b \beta})^{\dag}\;,\;\;\lambda_{\alpha}^{A} = \epsilon_{\alpha \beta}\epsilon^{AB}(\lambda_{\beta}^{B})^{\dag} \;,
\end{equation}
\begin{equation}
	\bar\lambda_{a}^{\dot\alpha} = -\epsilon^{\dot\alpha \dot\beta}\epsilon_{a b}(\bar\lambda_{b}^{\dot\beta})^{\dag}\;,\;\;\bar\lambda^{A \dot{\alpha}} = \epsilon^{\dot\alpha \dot\beta}\epsilon^{AB}(\lambda^{B\dot{\beta}})^{\dag} \;.
\end{equation}

\end{document}